\documentclass[pre,aps,twocolumn]{revtex4}%
\usepackage{times}
\usepackage{amsmath}
\usepackage{graphicx}
\usepackage{amssymb}
\usepackage{color}
\usepackage[normalem]{ulem}
\usepackage{amsfonts}%
\pdfoutput=1
\newcommand{\wb}{\omega_\mathrm{b}}

\begin{document}
\title{Impulse-induced generation of stationary and moving discrete breathers in
nonlinear oscillator networks}
\author{J. Cuevas--Maraver \thanks{Email: jcuevas@us.es}}
\affiliation{Grupo de F\'{\i}sica No Lineal, Departamento de F\'{\i}sica Aplicada I,
Universidad de Sevilla. Escuela Polit{\'e}cnica Superior, C/ Virgen de
\'Africa, 7, E-41011 Sevilla, Spain}
\affiliation{Instituto de Matem\'aticas de la Universidad de Sevilla (IMUS). Edificio
Celestino Mutis. Avda. Reina Mercedes s/n, E-41012 Sevilla, Spain}
\author{R. Chac\'{o}n \thanks{Email: rchacon@unex.es}}
\affiliation{Departamento de F\'{\i}sica Aplicada, E.I.I., Universidad de Extremadura,
Apartado Postal 382, E-06006 Badajoz, Spain }
\affiliation{Instituto de Computaci\'{o}n Cient\'{\i}fica Avanzada (ICCAEx), Universidad de
Extremadura, E-06006 Badajoz, Spain}
\author{F. Palmero \thanks{Email: palmero@us.es}}
\affiliation{Grupo de F\'{\i}sica No Lineal, Departamento de F\'{\i}sica Aplicada I,
Universidad de Sevilla. Escuela T\'ecnica Superior de Ingener\'{\i}a
Inform\'atica. Avda. Reina Mercedes s/n, E-41012 Sevilla, Spain}

\begin{abstract}
We study discrete breathers in prototypical nonlinear oscillator networks
subjected to non-harmonic zero-mean periodic excitations. We show how the
generation of stationary and moving discrete breathers are optimally
controlled by solely varying the impulse transmitted by the periodic
excitations, while keeping constant the excitation's amplitude and period. Our
theoretical and numerical results show that the enhancer effect of increasing
values of the excitation's impulse, in the sense of facilitating the
generation of stationary and moving breathers, is due to a correlative
increase of the breather's action and energy.

\end{abstract}
\maketitle


\section{Introduction}

\label{sec:intro}

Discrete breathers are intrinsic localized modes that can emerge in networks
of coupled nonlinear oscillators \cite{breather_reviews,Aubry}. They have been
observed not only in Hamiltonian lattices but also in driven dissipative
systems under certain conditions. Discrete breathers have been theoretically
predicted or experimentally generated in a wide variety of physical systems
such as Josephson junction arrays \cite{JJ}, coupled pendula chains
\cite{pendula}, micro- and macro-mechanical cantilever arrays
\cite{cantilever}, granular crystals \cite{granular}, nonlinear electrical
lattices \cite{electric}, and double-strand DNA models \cite{DNA}, just to
cite a few instances.

Up to now, breathers have been mainly studied for the case of a harmonic
external excitation, while various types of periodic excitations are in
principle possible, depending upon the physical context under consideration.
Since there are infinitely many different wave forms, a quite natural question
is to ask how the generation and dynamics of breathers are affected by the
presence of a generic periodic excitation.

In this present work, we show that a relevant quantity properly characterizing
the effectiveness of zero-mean periodic excitations $F(t)$ having equidistant
zeros at controlling the generation and dynamics of discrete breathers is the
\textit{impulse} transmitted by the external excitation over a half-period
(hereafter referred to simply as the excitation's impulse \cite{1},
$I\equiv\int_{0}^{T/2}F(t)dt$, $T$ being the period)$-$ a quantity integrating
the conjoint effects of the excitation's amplitude, period, and waveform. It
is worth mentioning that the relevance of the excitation's impulse has been
observed previously in quite different contexts, such as ratchet transport
\cite{2}, adiabatically ac driven periodic (Hamiltonian) systems \cite{3},
driven two-level systems and periodically curved waveguide arrays \cite{4},
chaotic dynamics of a pump-modulation Nd:YVO$_{4}$ laser \cite{5}, topological
amplification effects in scale-free networks of signaling devices \cite{6},
and controlling chaos in starlike networks of dissipative nonlinear
oscillators \cite{7}.

The rest of this paper is organized as follows. In Sec. II we introduce the
model and further comment on some of its main features. Section III provides
numerical evidence for the essential role of the excitation's impulse at
generating breathers and controlling their stability for the prototypical
cases of a hard $\phi^{4}$ potential and a sine-Gordon potential. A
theoretical explanation of the effectiveness of the excitation's impulse in
terms of energy and action is provided in Sec. IV. Finally, Sec. V is devoted
to a discussion of the major findings and of some open problems.

\section{Model system}

\label{sec:model}

\begin{figure}[tb]
\begin{center}
\begin{tabular}[c]{cc}
    \includegraphics[width=4cm]{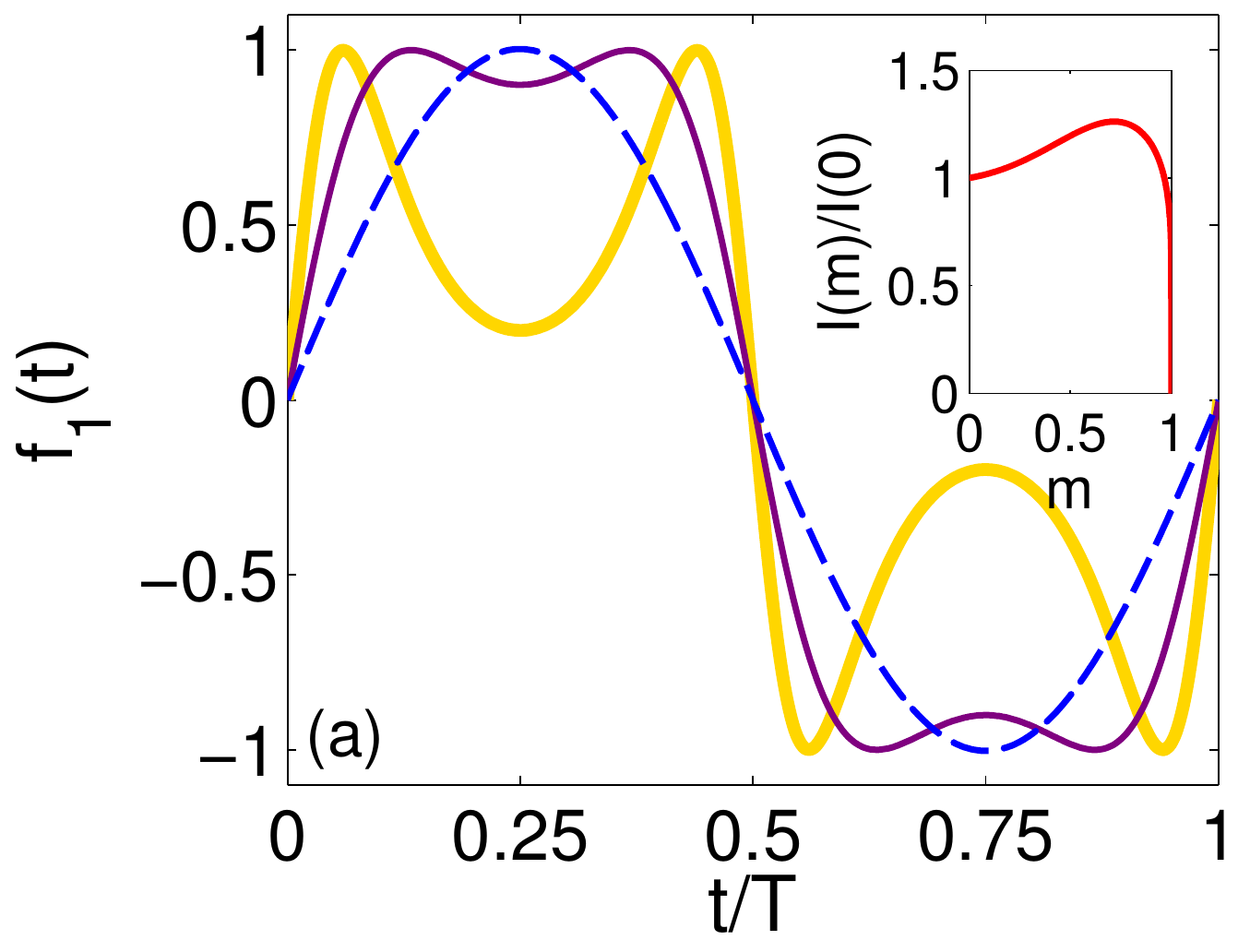}
    \includegraphics[width=4cm]{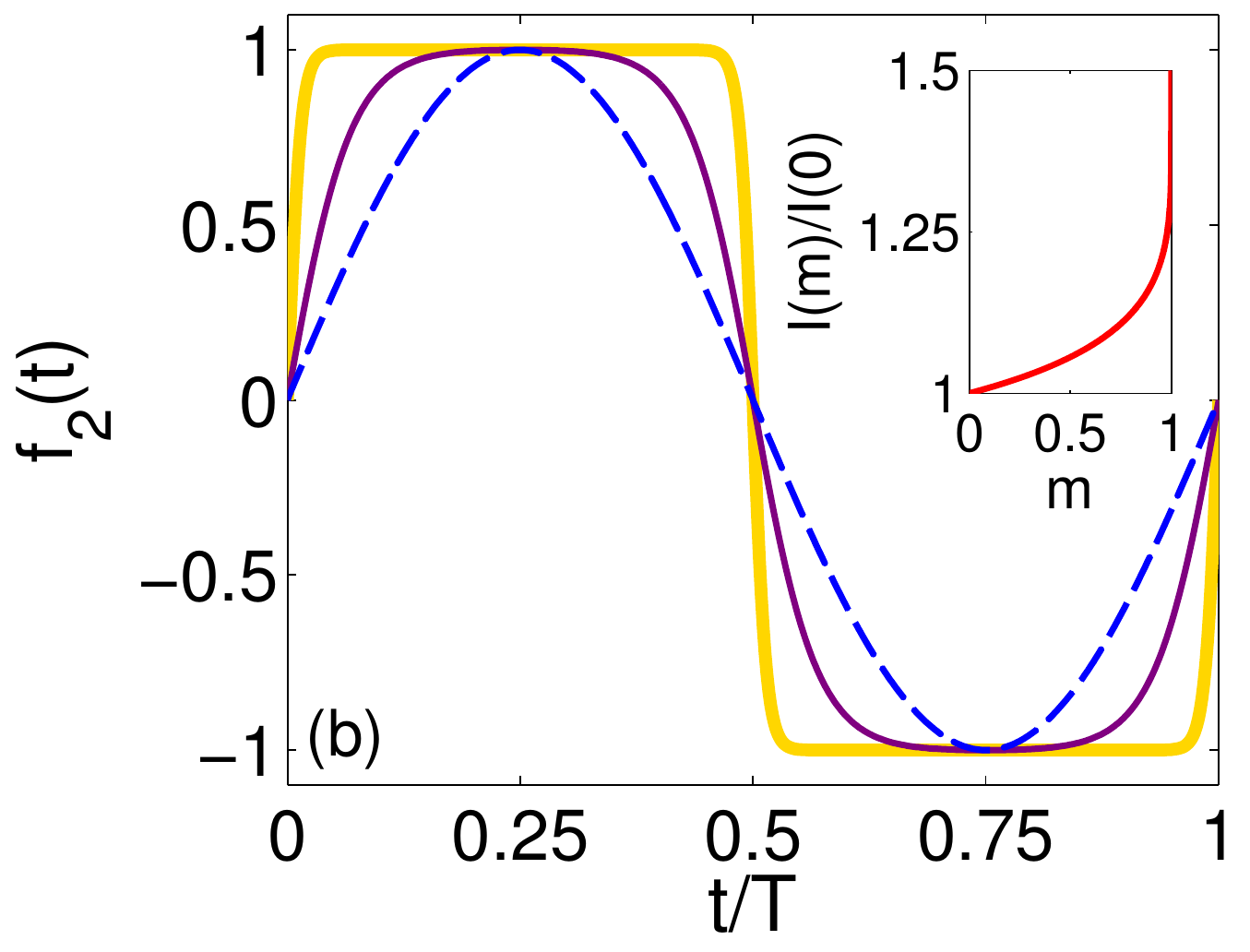}
\end{tabular}
\end{center}
\caption{External excitations (\ref{eq:driving1}) and
(\ref{eq:driving2}) vs time over a period (panels (a) and (b),
respectively). In the former case, dashed (blue), thin (purple), and thick
(golden) lines correspond to $m=0$, $m=m_{\max}=0.717$, and $m=0.99$,
respectively. In the latter case, the same lines correspond to $m=0$,
$m=0.99,$ and $m=1-10^{-14}$, respectively. Notice that, when $m\rightarrow1$,
the excitation $f_{2}$ becomes a square wave signal while $f_{1}$ vanishes.
The insets show the respective normalized impulses of the excitations vs the
shape parameter (see the text).}%
\label{fig:driving}%
\end{figure}

The discrete nonlinear Klein--Gordon equation with linear coupling, linear
damping, and external periodic excitation is one of the simplest equations
where dissipative discrete breathers may arise:%

\begin{equation}
\ddot{u}_{n}+\alpha\dot{u}_{n}+V^{\prime}(u_{n})+C(2u_{n}-u_{n+1}%
-u_{n-1})=F(t), \label{eq:dyn}%
\end{equation}
in which $V(u_{n})$ is an on--site (substrate) potential, $\alpha$ is the
damping constant, $C$ is the coupling constant, while $F(t)$ is a zero-mean
periodic excitation.
\begin{equation}
F(t)=(-1)^{hn}f_{0}f_{1,2}(t), \label{eq:driving}%
\end{equation}
{ }in which $f_{0}$ is the driving amplitude, $h$ is a hardness parameter
whose value is $0$ ($1$) when the on-site potential is soft (hard), while
$f_{1}(t)$, $f_{2}(t)$ are two different periodic excitations that we will use
as illustrative examples to show that the impulse is the relevant quantity
controlling the effect of the external excitation on the generation and
stability properties of breathers. These periodic excitations are given by
\begin{equation}
f_{1}(t)=N(m)\mathrm{sn}\left(  \frac{2K(m)\omega_{\mathrm{b}}t}{\pi
};m\right)  \mathrm{dn}\left(  \frac{2K(m)\omega_{\mathrm{b}}t}{\pi};m\right)
, \label{eq:driving1}%
\end{equation}%
\begin{equation}
f_{2}(t)=\mathrm{sn}\left(  \frac{2K(m)\omega_{\mathrm{b}}t}{\pi};m\right)  ,
\label{eq:driving2}%
\end{equation}
where $\mathrm{sn}(\cdot)\equiv\mathrm{sn}(\cdot;m)$ and $\mathrm{dn}%
(\cdot)\equiv\mathrm{dn}(\cdot;m)$ are Jacobian elliptic functions of
parameter $m$ [$K\equiv K(m)$ is the complete elliptic integral of the first
kind], while $N(m)$ is a normalization function which is introduced for the
elliptic excitation $f_{1}(t)$ to have the same amplitude $f_{0}$ and period
$T\equiv2\pi/\omega_{\mathrm{b}}$ for any wave form (i.e., $\forall
m\in\left[  0,1\right)  $ ). Specifically, the normalization function is given
by
\begin{equation}
N(m)=\left[  \eta_{1}+\frac{\eta_{2}}{1-\exp[(m-\eta_{3})/\eta_{4}]}\right]
^{-1},
\end{equation}
with $\eta_{1}=0.43932$, $\eta_{2}=0.69796$, $\eta_{3}=0.37270$, $\eta
_{4}=0.26883$. In both excitations $f_{1,2}(t)$, when $m=0$, then
$F(t)=(-1)^{hn}f_{0}\sin\left(  \omega_{\mathrm{b}}t\right)  $, that is, one
recovers the standard case of a harmonic excitation \cite{Cubero}, whereas,
for the limiting value $m=1$, the excitation $f_{1}(t)$ vanishes while the
excitation $f_{2}(t)$ reduces to a square wave. It is worth noting that the
excitations $f_{1,2}(t)$ have been chosen to exhibit the following properties.
For the excitation $f_{1}(t)$, its impulse per unit of amplitude, $I_{1}(m)/I_{1}(0)$ with
$I_{1}(m)=TN(m)/\left[2 K(m)\right]  $, presents a single maximum at
$m=m_{\max}\simeq0.717$. For the excitation $f_{2}(t)$, its impulse is written
$I_{2}(m)=T\mathrm{arctanh}(\sqrt{m})/\left[2\sqrt{m}K(m)\right]$, and hence it
corresponding normalized impulse $I_{2}(m)/I_{2}(0)$ grows monotonically
from $1$ to $1.5$. Figure \ref{fig:driving} shows the time-dependence of both
excitations over a period together with the dependence of their respective
normalized impulses on the shape parameter $m$.

The use of Jabobian elliptic functions as periodic excitations is mainly motivated by the fact that, after normalizing their (natural) arguments to keep the period as a fixed independent parameter, their waveforms can be changed by solely varying a {\em single} parameter: the elliptic parameter $m$, and hence the corresponding impulse will only depend on $m$ once the amplitude and period are fixed.

The aim of this paper is to study the effectiveness of the excitation's
impulse at controlling breathers arising in Eq. (\ref{eq:dyn}) by considering
two prototypical on--site potentials. First, a hard $\phi^{4}$ potential,
which was previously considered in \cite{Cubero} for the limiting case of a
harmonic excitation $\left(  m=0\right)  $, and where it was shown that there
exists a threshold value of $f_{0}$ such that breathers do not exist below it.
Remarkably, such a threshold amplitude can be decreased in the presence of
noise through a stochastic resonance mechanism. Second, a sine-Gordon
potential \cite{Marin} so that Eq. (\ref{eq:dyn}) becomes the so-called
Frenkel--Kontorova model (see, e.g., Refs. \cite{Kivshar,sG} for additional
details), in which the emergence of discrete moving breathers \cite{moving} is
indicated by the existence of a pitchfork bifurcation together with the
appearance of an intermediate state.

The existence of discrete breathers is characterized by using techniques based
on the anti-continuous (AC) limit \cite{AClimit}. Thus, two periodic
attractors must be found in such a limit (i.e., for the corresponding isolated
nonlinear oscillator) such that the attractor with the largest amplitude is
assigned to the central ($n=0$) site of the chain, while the other periodic
attractor is assigned to the rest of the coupled oscillators. Such a solution
is then continued from $C=0$ to the prescribed value of $C$. Since discrete
breathers are periodic orbits in phase space, they can be calculated by means
of a shooting method, i.e., they can be considered as fixed points of the map:%

\begin{equation}
\left[  \{u_{n}(0)\},\{\dot{u}_{n}(0)\}\right]  \rightarrow\left[
\{u_{n}(T)\},\{\dot{u}_{n}(T)\}\right]  .
\end{equation}

This analysis is
accomplished by using a Powell hybrid algorithm complemented by an 5th-6th
order Runge--Kutta--Verner integrator. To study the stability of discrete
breathers, a small perturbation $\xi_{n}$ is introduced to a given $u_{n0}$
solution of Eq. (\ref{eq:dyn}) according to $u_{n}=u_{n0}+\xi_{n}$. Thus, one
obtains the equation which is verified (to first order) by $\xi_{n}$:
\begin{equation}
\ddot{\xi}_{n}+\alpha\dot{\xi}_{n}+V^{\prime\prime}(u_{n0})\xi_{n}+C(2\xi
_{n}-\xi_{n+1}-\xi_{n-1})=0.
\end{equation}
To determine the orbital stability of periodic orbits, a Floquet analysis can
be performed so that the stability properties are deduced from the spectrum of
the Floquet operator (whose matrix representation is the monodromy
$\mathcal{M}$), given by
\begin{equation}
\left[  \{\xi_{n}(T)\},\{\dot{\xi}_{n}(0)\}\right]  ^{T}=\mathcal{M}\left[
\{\xi_{n}(0)\},\{\dot{\xi}_{n}(0)\}\right]  ^{T},
\end{equation}
where $\lambda=\exp(i\theta)$ are the \textit{Floquet multipliers} while the
values of $\theta$ are the \textit{Floquet exponents}. All eigenvalues must
lie inside the unit circle if the breather is stable.

\section{Numerical results}

\label{sec:numerics}

\begin{figure}[tb]
\begin{center}%
\begin{tabular}
[c]{cc}%
\includegraphics[width=4cm]{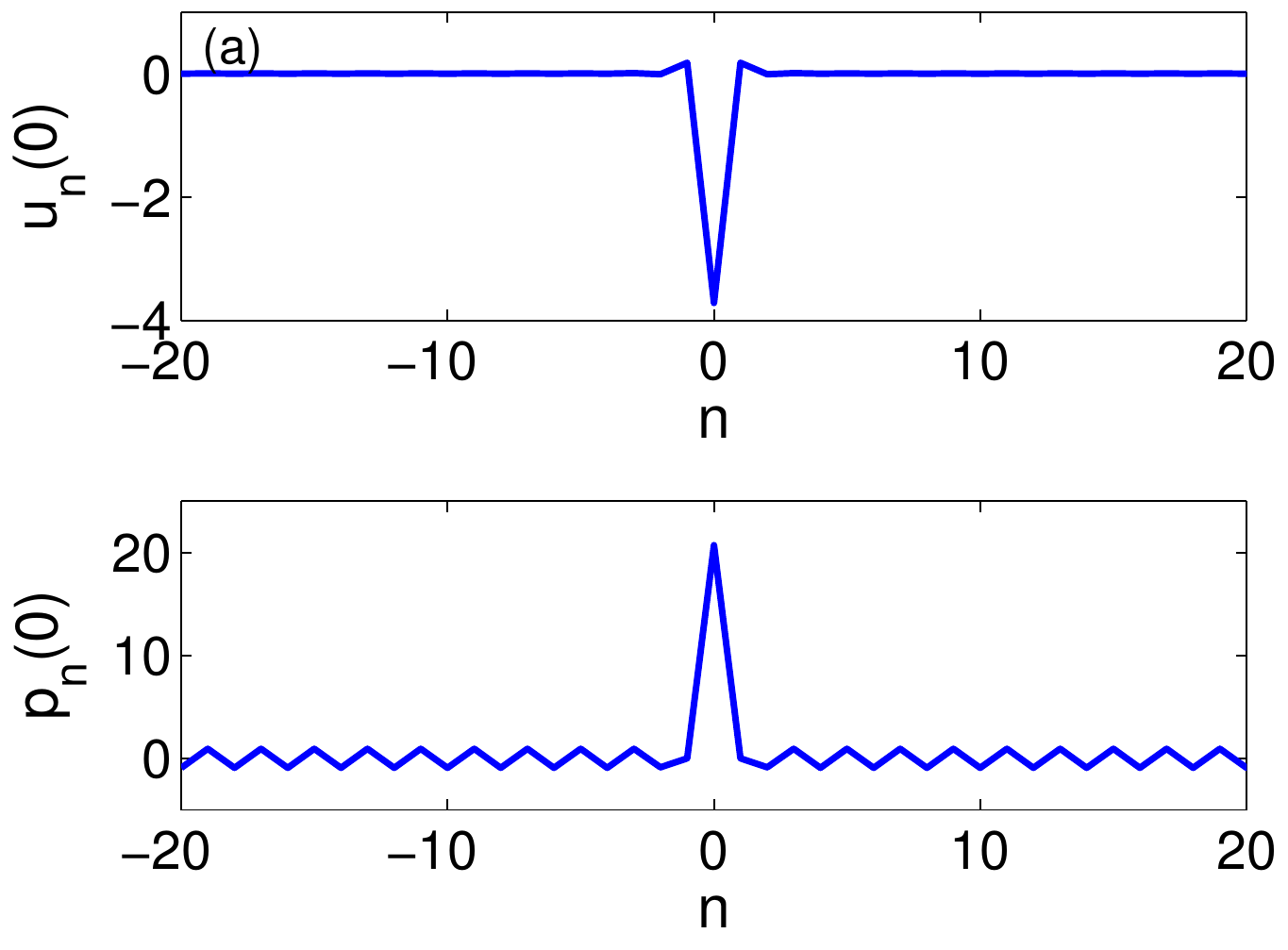} &
\includegraphics[width=4cm]{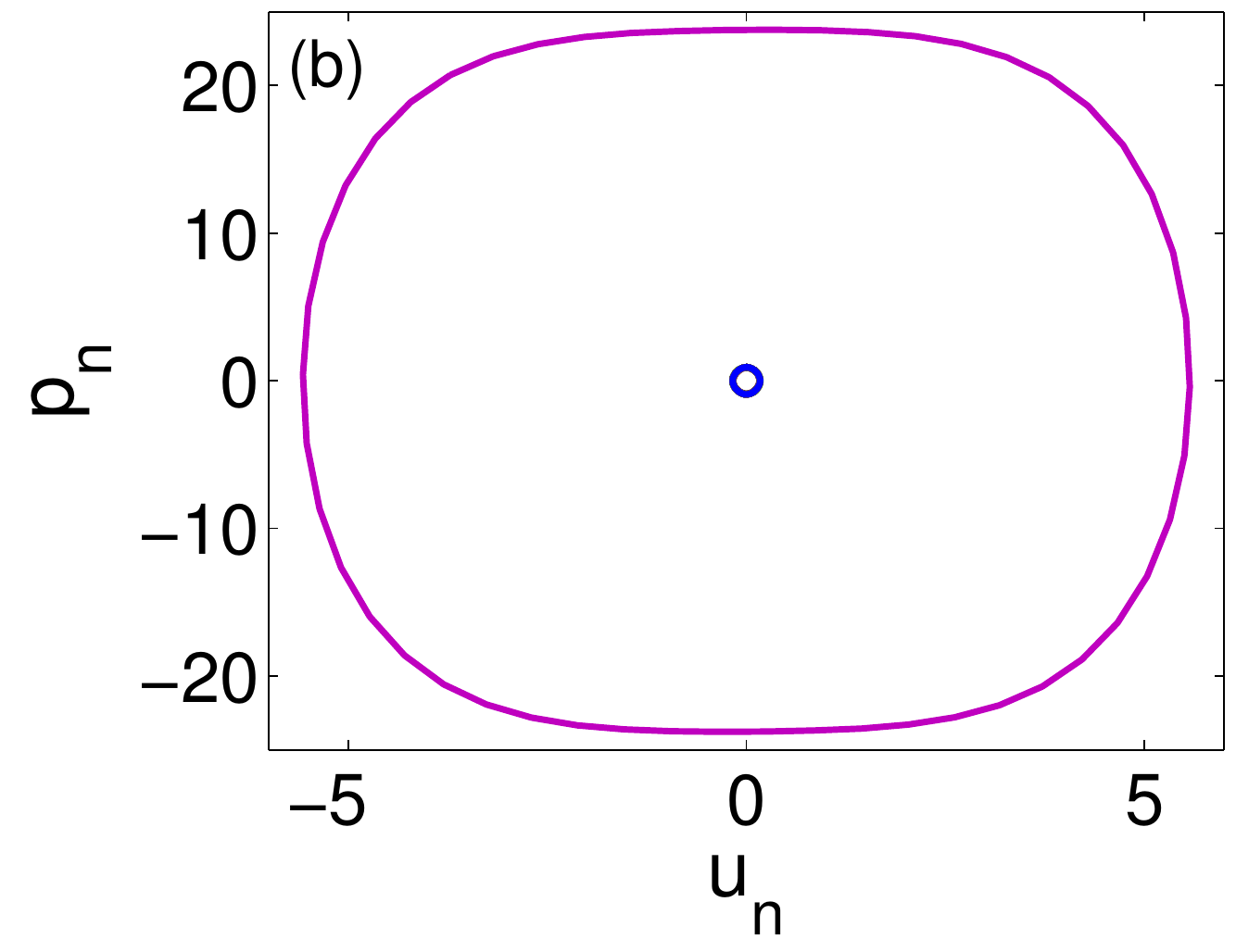}\\
\includegraphics[width=4cm]{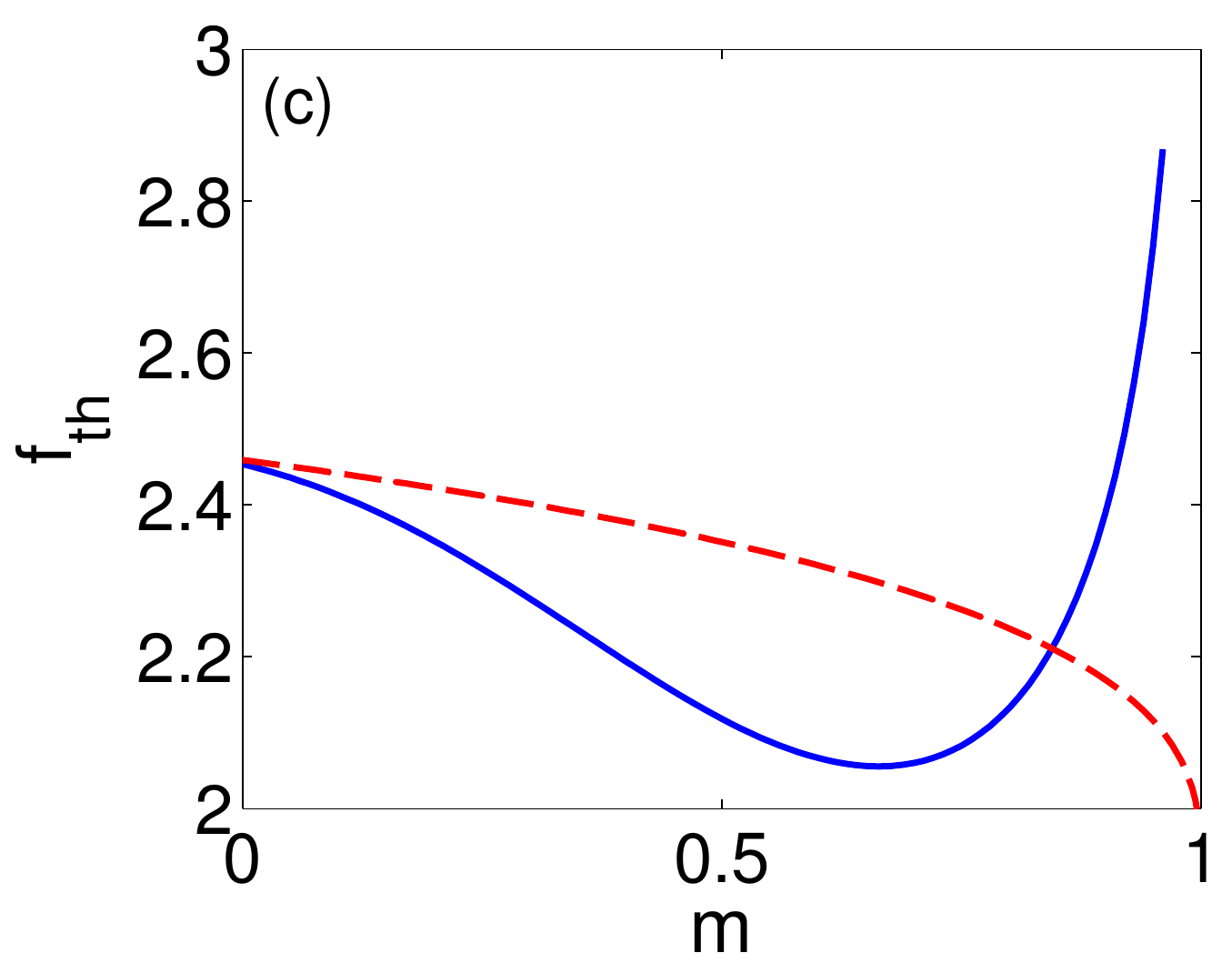} &
\includegraphics[width=4cm]{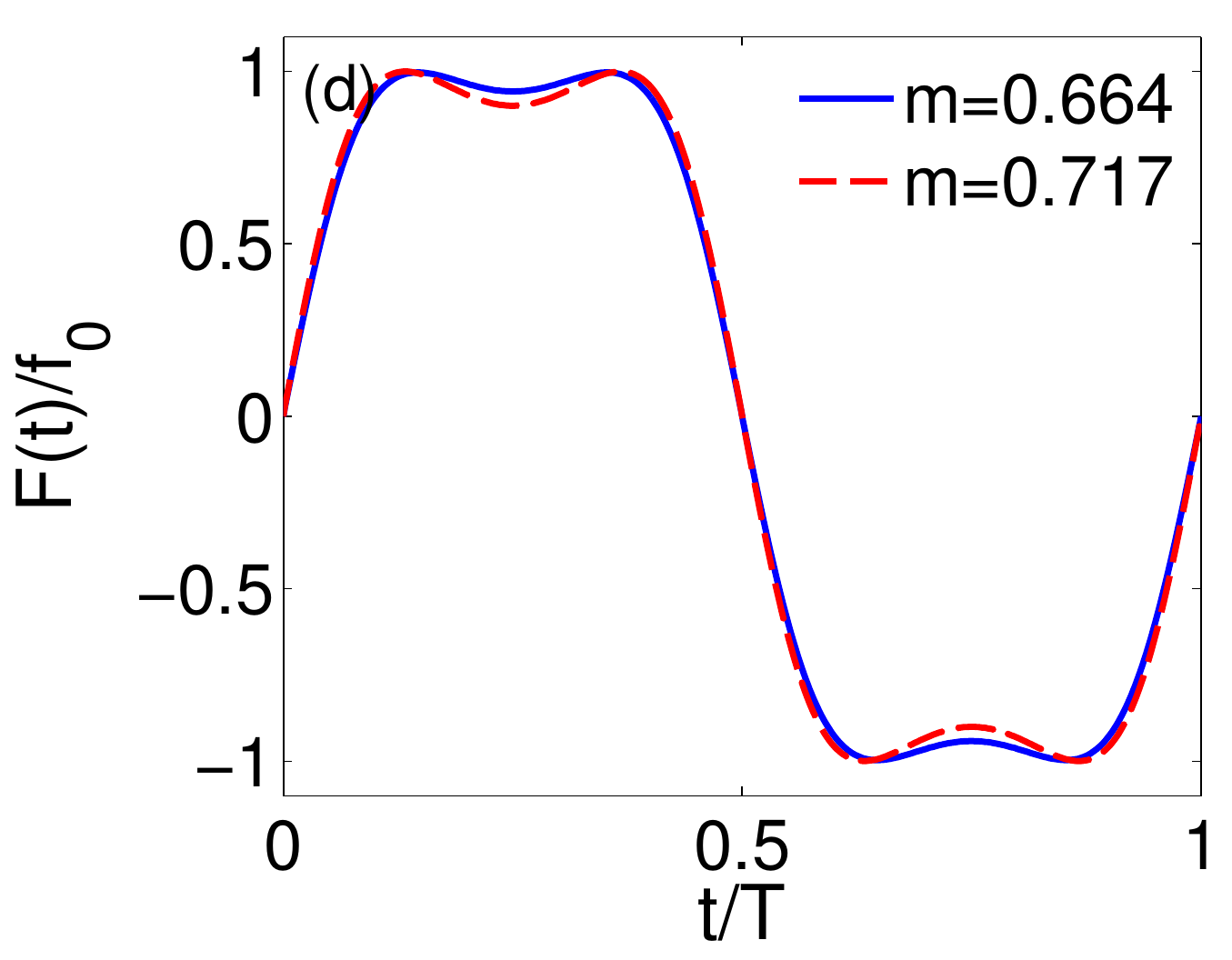}\\
&
\end{tabular}
\end{center}
\caption{Profiles of $u_{n},p_{n}\equiv\dot{u}_{n}$ (a) and phase space
diagrams (b) of a breather in a $\phi^{4}$ chain for $m=0.6$ and $f_{0}=3$.
(c) Threshold amplitude for the existence of breathers vs shape parameter $m$.
Solid (blue) and dashed (red) lines correspond to excitations
(\ref{eq:driving1}) and (\ref{eq:driving2}), respectively. Notice that in the
former case, there exists a minimum at $m=m_{c}=0.664$ for $f_{\mathrm{th}%
}=2.056$, whereas in the latter case, one sees a monotonously decreasing
behavior. (d) External excitation (\ref{eq:driving1}) vs time over a period
for two values of the shape parameter: $m=m_{c}$, for which $f_{\mathrm{th}}$
is minimum, and $m=m_{\max}=0.717$, for which the impulse is maximum. Fixed
parameters: $\alpha=0.1$, $C=1$, $\omega_{\mathrm{b}}=5$.}%
\label{fig:hard}%
\end{figure}

Our numerical study starts with the case of a hard $\phi^{4}$ potential, that
is, $V(u)=u^{2}/2+u^{4}/4$. Notice that breathers in such a potential exhibit
staggered tails due to its hardness. This means that the system must be driven
following this pattern by taking $h=1$ in (\ref{eq:driving}) (see Figs.
\ref{fig:hard}(a) and \ref{fig:hard}(b)). It has been shown for $m=0$
\cite{Cubero} that breathers exist if $f_{0}>f_{\mathrm{th}}$, i.e., the
excitation amplitude must surpass a certain threshold. In general, this
threshold is a function of the system parameters. Here we study the dependence
of this threshold on the shape parameter $m$, while keeping fixed the
remaining parameters. Figure \ref{fig:hard}(c) shows this dependence for the
parameters $\alpha=0.1$, $C=1$, $\omega_{\mathrm{b}}=5$.

For the excitation $f_{1}(t)${\textbf{ }}(\ref{eq:driving1}), it should be
emphasized the existence of a minimum threshold at a critical value
$m=m_{c}\approx0.664$ for such set of parameters; however, if any of such
parameters were varied, $m_{c}$ would remain close to such a value. Although
this critical value does not exactly match the value $m=$ $m_{\max}%
\approx0.717$ at which the impulse $I_{1}\left(  m\right)  $ presents a single
maximum, it is very close to the value $m=m_{1}=0.6416$ where the first
harmonic of the Fourier expansion of the external driving presents a single
maximum. Also, the waveforms corresponding to $m_{c}$ and $m_{\max}$ can be
hardly distinguishable, as is shown in Fig. \ref{fig:hard}(d), which means
that the values of their respective impulses are almost identical (the
relative difference is only $\sim0.42\%$). The fact that $m_{c}$ does not
change significantly when $\omega_{\mathrm{b}}$ and $C$ are varied implies
that this property holds in the AC limit. Indeed, for the isolated oscillator
we found that for the largest-amplitude attractor there exists a minimum value
of its amplitude, $f_{\mathrm{th}}$, at $m=0.668\approx m_{c}$, while the
smallest-amplitude attractor exists for any value of $f_{0}$ (i.e.,
$f_{\mathrm{th}}=0,\ \forall m$). Thus, the breather seems to inherit this key
feature (impulse-induced threshold behaviour) of the largest-amplitude
attractor of the isolated oscillator. It is worth mentioning that $\alpha$
must be sufficiently small in order that two periodic attractors can exist in
the AC limit.

For the excitation $f_{2}(t)$ (\ref{eq:driving2}), we found that the threshold
amplitude $f_{\mathrm{th}}$ exhibits a monotonously decreasing behavior as a
function of the shape parameter (see \ref{fig:hard}(c)), as expected from the
monotonously increasing behavior of its impulse $I_{2}\left(  m\right)  $.
Thus, the analysis of both periodic excitations $f_{1,2}(t)$ confirmed the
same effect of the excitation's impulse on the amplitude threshold for the
existence of breathers.

\begin{figure}[tb]
\begin{center}%
\begin{tabular}
[c]{cc}%
\includegraphics[width=4cm]{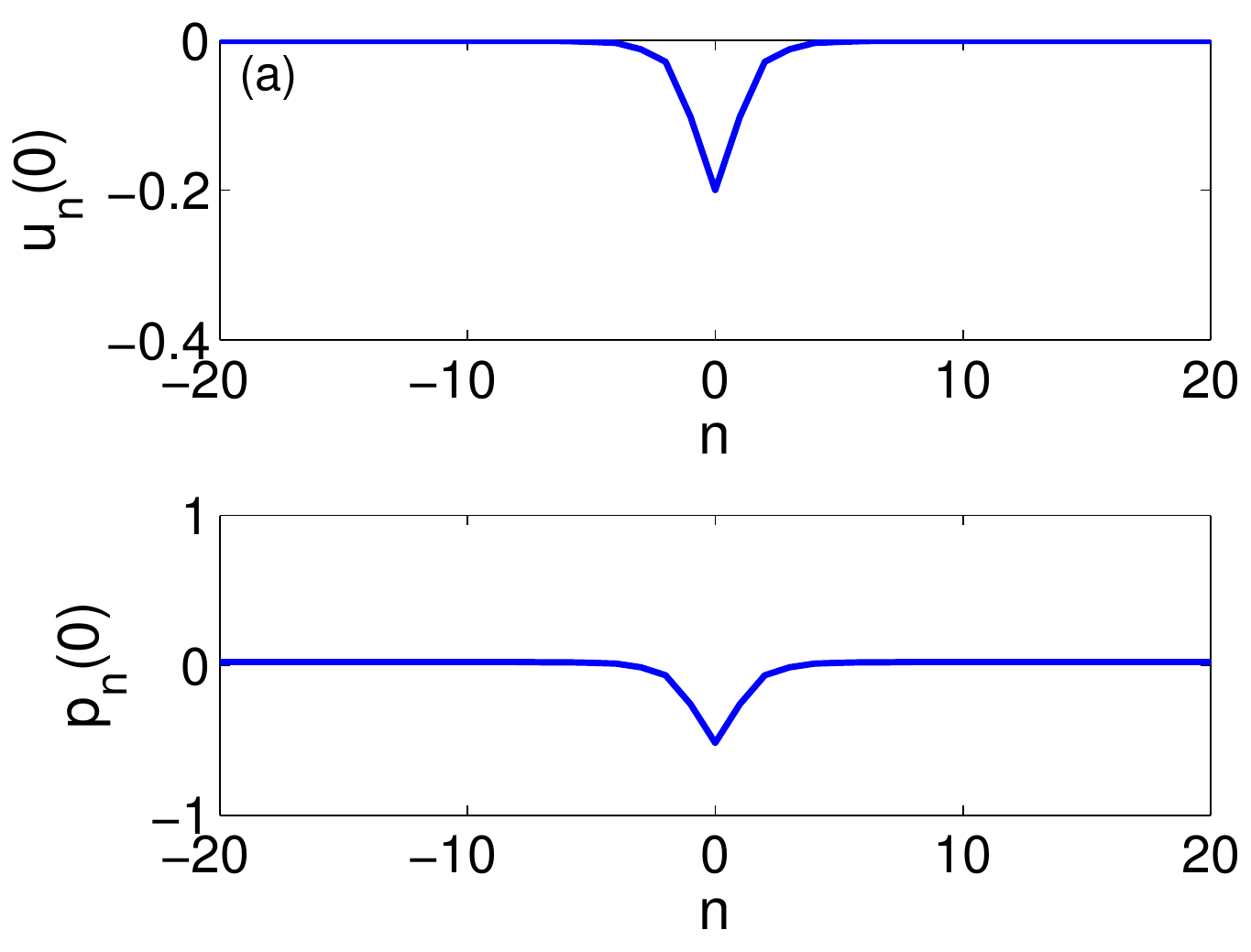} &
\includegraphics[width=4cm]{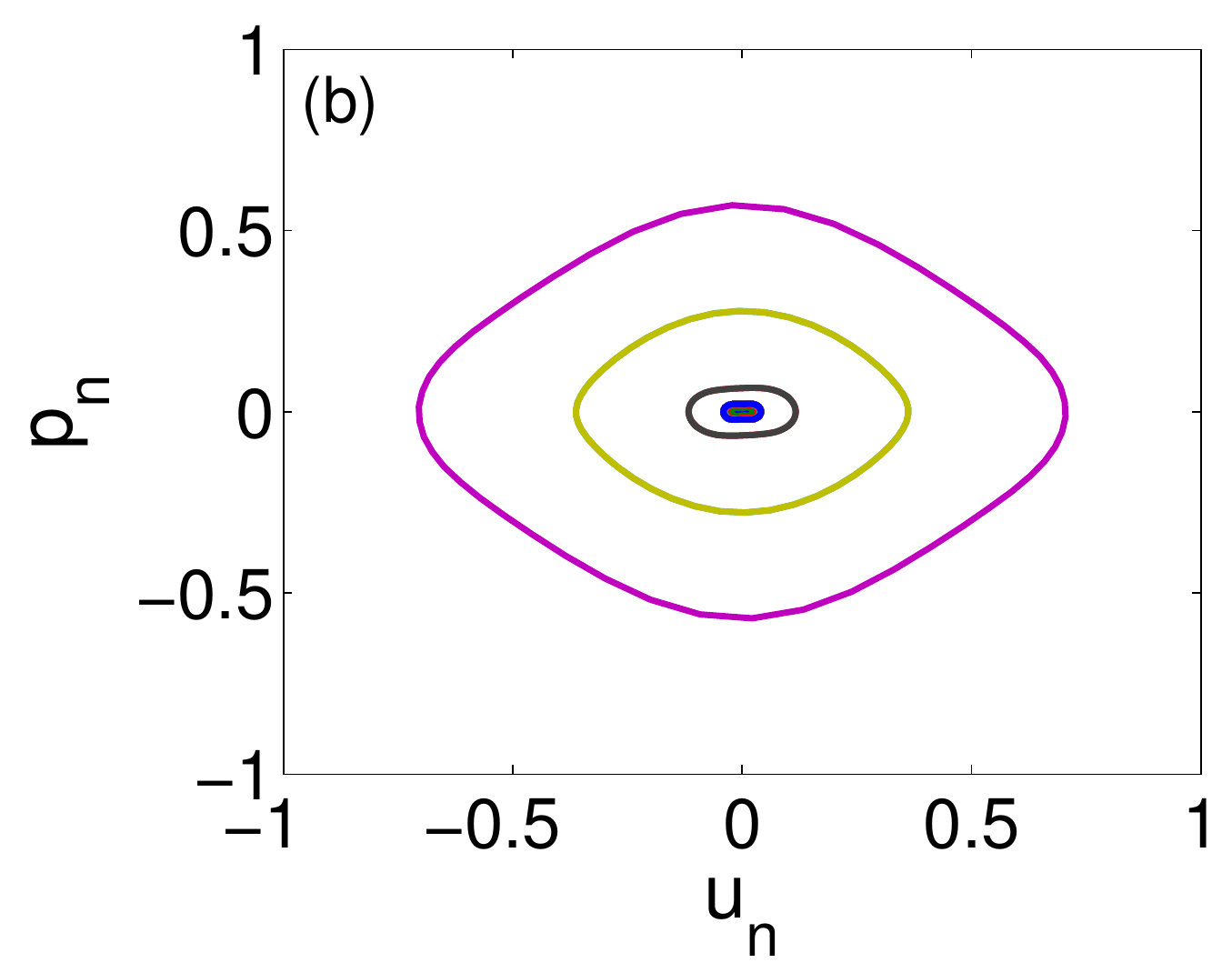}\\
\includegraphics[width=4cm]{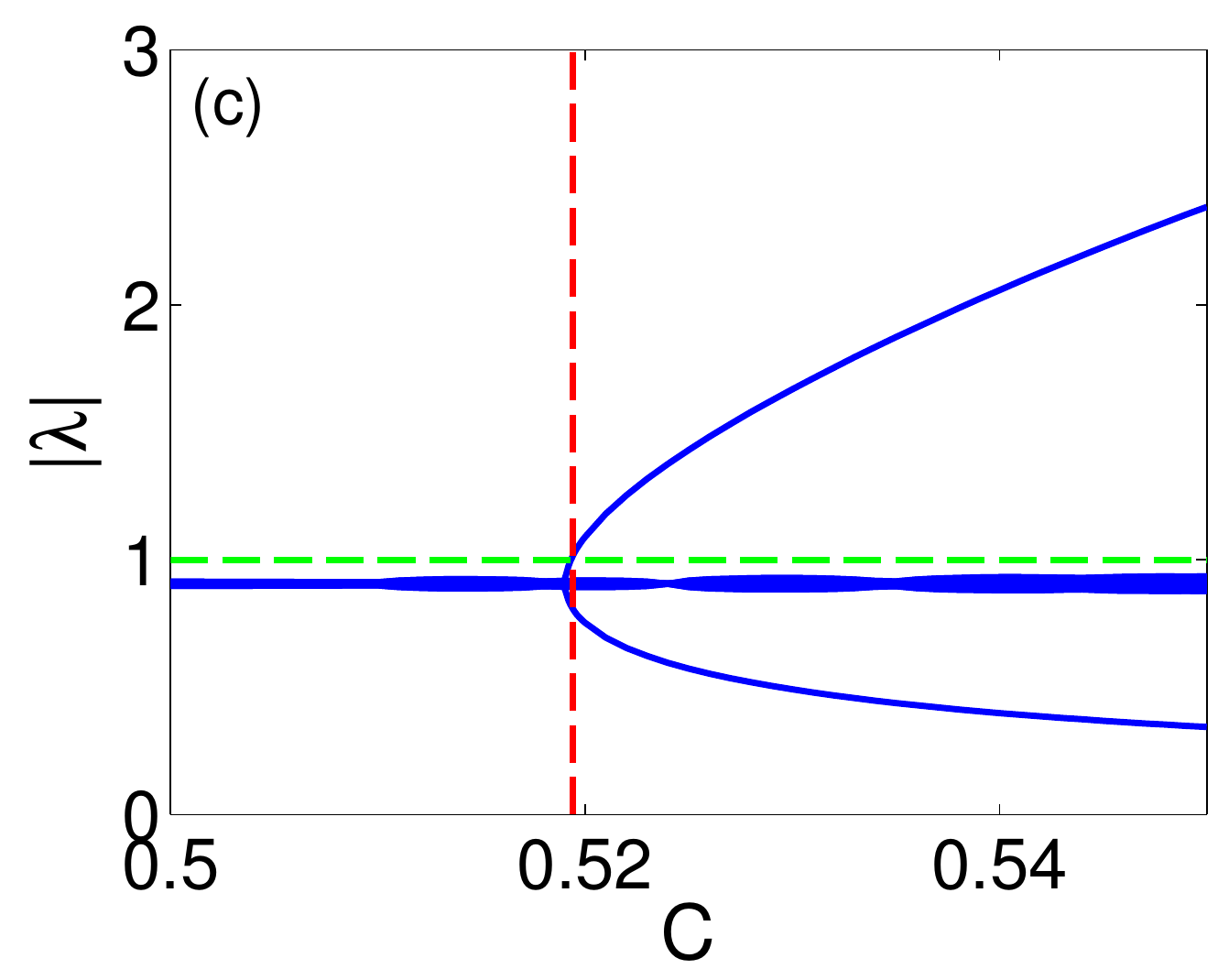} &
\includegraphics[width=4cm]{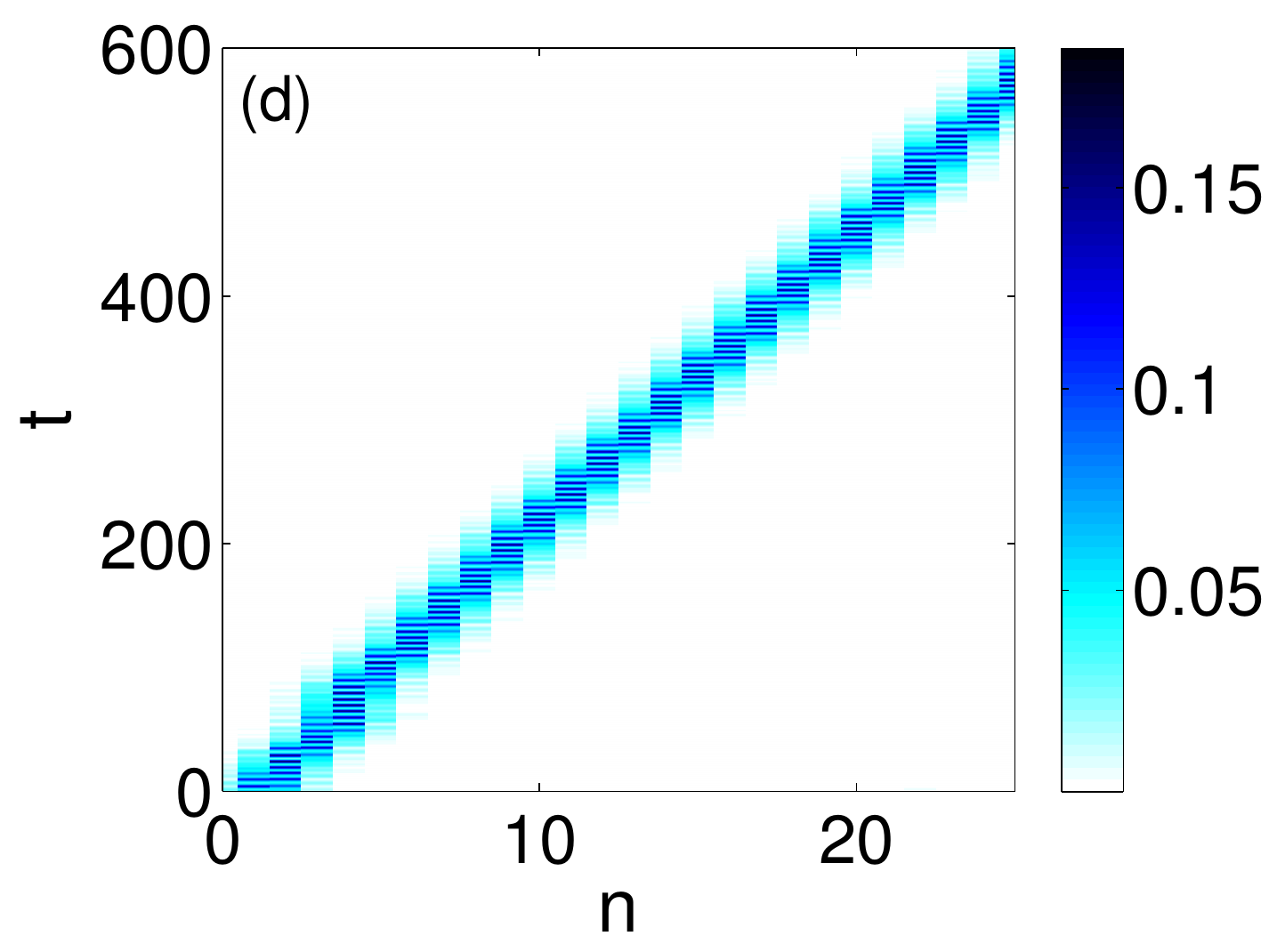}\\
&
\end{tabular}
\end{center}
\caption{Profiles of $u_{n},p_{n}\equiv\dot{u}_{n}$ (a) and phase space
diagrams (b) of a breather in a chain with a sine-Gordon potential for the
excitation (3) and $C=0.5$. (c) Modulus of the Floquet multiplier $\lambda$ vs
coupling $C$. The dashed vertical line indicates de location of $C_{1}$. (d)
Spatiotemporal pattern of the energy density of a moving breather emerging
from the unstable solution for $C=0.52$. Fixed parameters: $\alpha=f_{0}%
=0.02$, $m=0.66$, $\omega_{\mathrm{b}}=0.2\pi$.}%
\label{fig:soft}%
\end{figure}

\begin{figure}[tb]
\begin{center}%
\begin{tabular}
[c]{cc}%
\includegraphics[width=4cm]{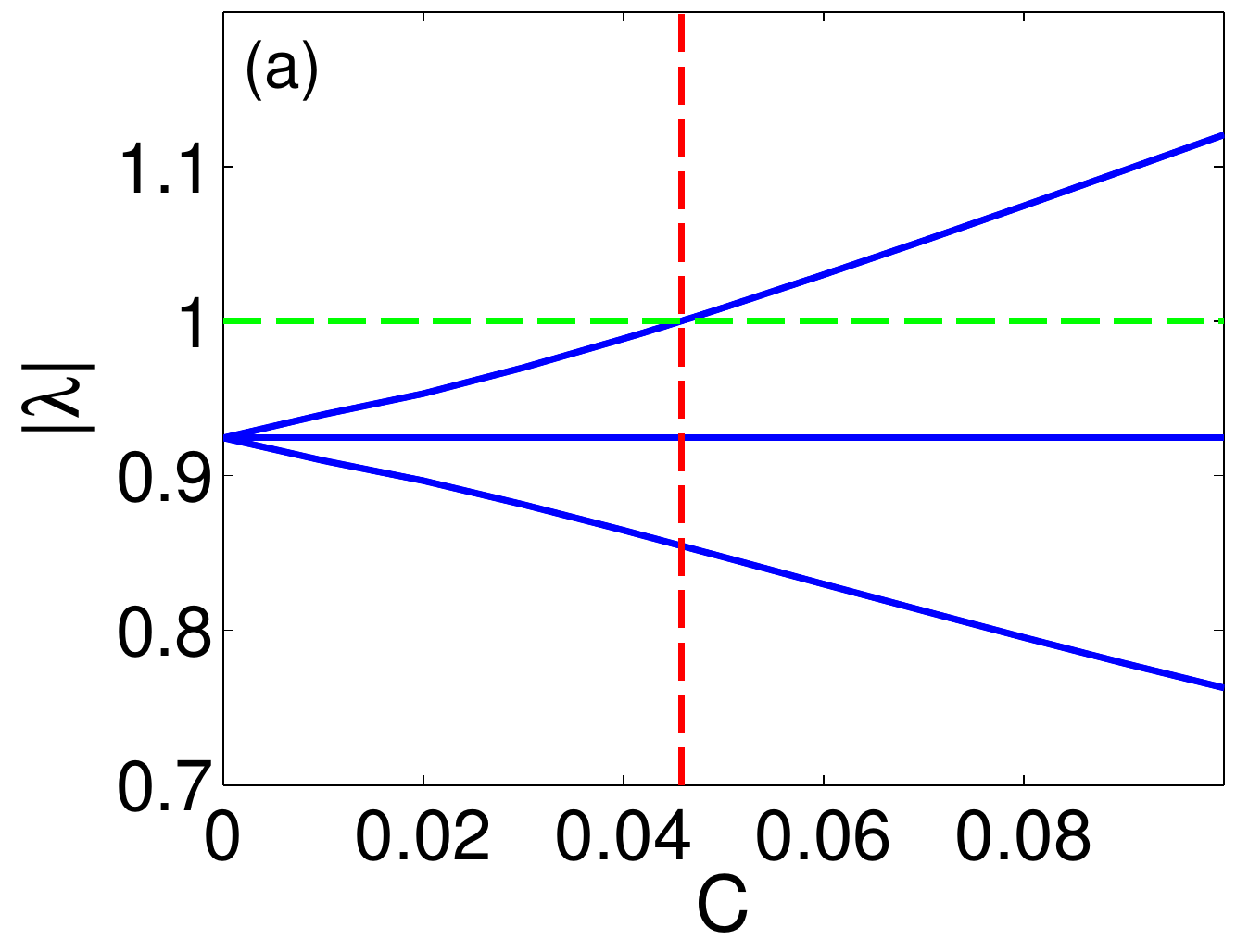} &
\includegraphics[width=4cm]{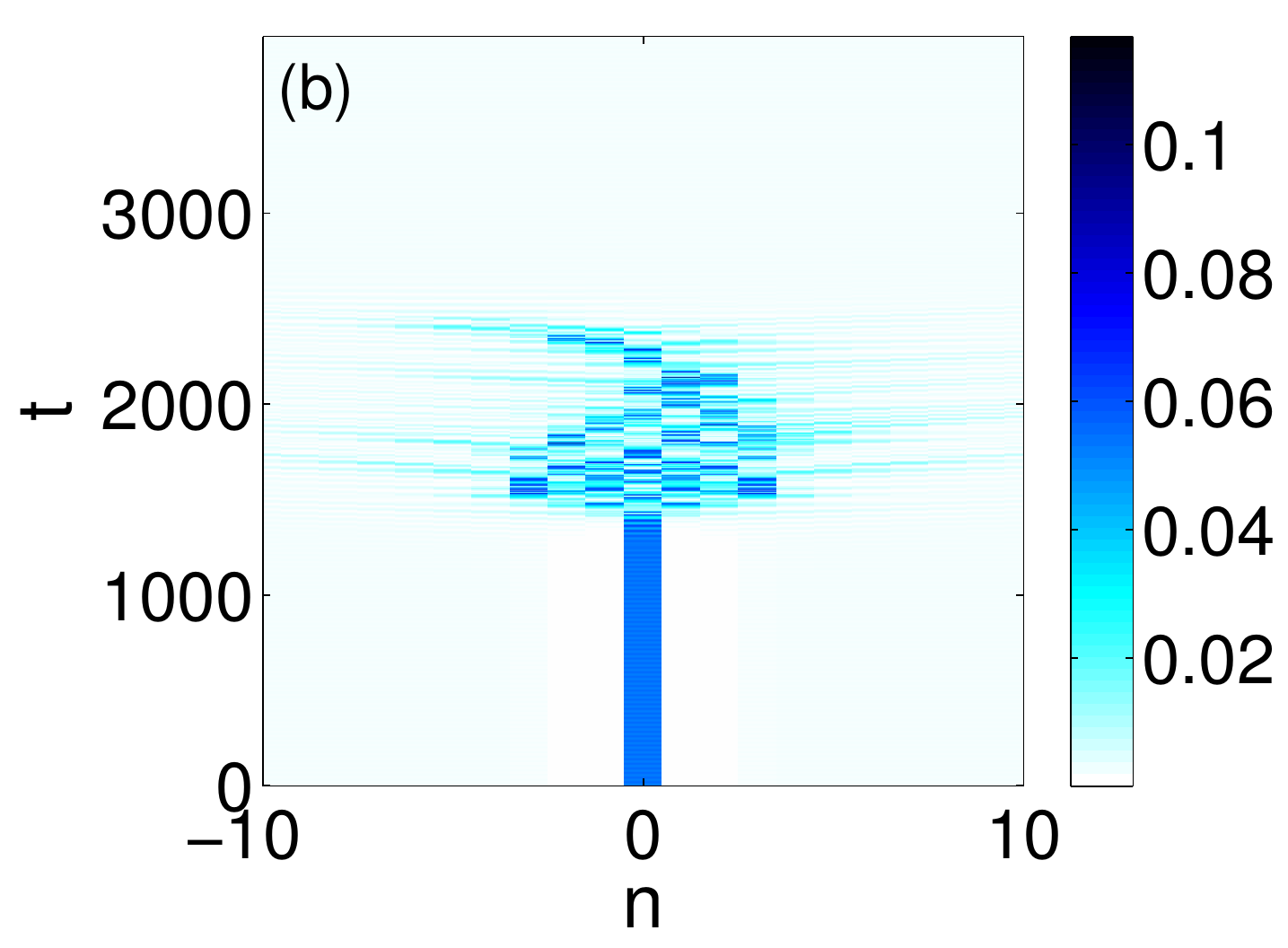}\\
&
\end{tabular}
\end{center}
\caption{(a) Modulus of the Floquet multiplier $\lambda$ vs coupling $C$ for
the excitation (3). The dashed vertical line indicates de location of $C_{2}$.
(b) Spatiotemporal pattern of the energy density of an unstable breather for
$C=0.1$. This instability finally leads to the destruction of the breather.
Fixed parameters: $m=0.66$, $\alpha=f_{0}=0.02$, $\omega_{\mathrm{b}}=0.8$.}%
\label{fig:soft2}%
\end{figure}

Next, we consider the Frenkel--Kontorova model, i.e., the case of a
sine-Gordon potential: $V^{\prime}(u)=\sin(2\pi u)/(2\pi)$. This case, which
was previously analyzed in detail for the limiting case of a harmonic
excitation $\left(  m=0\right)  $ in \cite{Marin}, is much richer than the
previous one due to the existence of exchange of stability bifurcations, Hopf
/ Neimark-Sacker bifurcations, chaos, moving breathers \cite{Meister}, and
rotobreathers \cite{roto}. Figure \ref{fig:soft} shows the position, velocity,
and phase space diagram of a typical breather. Since the sine-Gordon potential
is soft, tails are unstaggered and hence one takes $h=0$ in (\ref{eq:driving})
in order that the periodic excitation fit this pattern. According to Ref.
\cite{Marin}, for $\alpha=0.02$ and $f_{0}\gtrsim0.05$ the largest-amplitude
attractor in the AC limit corresponds to a rotation, and hence it cannot be
used for the analysis of breathers (the analysis of rotobreathers is beyond
the scope of the present work). Thus, we are fixing $\alpha=f_{0}=0.02$ in our
numerical simulations. Similarly to the case of a hard $\phi^{4}$ potential,
we found a threshold for the existence of breathers inheriting the features of
the largest-amplitude attractor of the AC limit \footnote{
The existence of such thresholds for the existence of breathers in
driven-damped systems seems to be a general phenomenon, as demonstrated in
\cite{Feckan} for localized periodic travelling waves in FPU lattices}.
Additionally, we found an interesting behaviour arising from stability
exchange bifurcations that leads to the onset of moving breathers
\cite{moving}. In this kind of bifurcations, a site-centered breather (i.e., a
breather with a single site excited at the AC limit) undergoes a supercritical
pitchfork bifurcation becoming unstable past a critical value of the coupling,
$C=C_{1},$ while a new kind of breather appears --- the so-called
\textit{intermediate breather} --- where two adjacent sites are excited with
different amplitudes. This intermediate breather disappears after undergoing a
subcritical pitchfork bifurcation at $C=C_{1}^{\prime}$. At this coupling
value, a site-centered breather (i.e., a breather with two adjacent sites
excited with the same amplitude), which is unstable for $C<C_{1}^{\prime}$,
changes its stability (see Fig. 22 in \cite{Aubry}, and Figs. 5 and 8 in
\cite{Marin}). Figures \ref{fig:soft}(c) and \ref{fig:soft}(d) show the onset
of this instability and the dynamics of a moving breather, respectively, for
the excitation $f_{1}(t)$. The spatiotemporal patterns of moving breathers is
illustrated by plotting their energy density:%

\begin{equation}
E_{n}=\frac{\dot{u}_{n}^{2}}{2}+V(u_{n})+\frac{C}{4}\left[  (u_{n}%
-u_{n+1})^{2}+(u_{n}-u_{n-1})^{2}\right]  .
\end{equation}
We found these results for frequencies over the range $1/2<\omega_{\mathrm{b}%
}<2/3$. Note, however, that over the range $2/3<\omega_{\mathrm{b}}<1$ the
discussed phenomenology can change due to the properties of Floquet exponents
\cite{stab}. Indeed, for frequencies over the range $2/3<\omega_{\mathrm{b}%
}<1$, the breather undergoes a Neimark--Sacker bifurcation as the coupling is
increased, making it unstable past a critical coupling value $C=C_{2}$. This
instability is characterized by the eventual destruction of the breather
(i.e., the localization is lost and only a linear mode remains; see Fig.
\ref{fig:soft2}). The critical value $C_{2}$ is much smaller than $C_{1}$ (in
fact, $C_{2}$ is close to 0, i.e. to the AC limit). Therefore, it has no sense
to study the emergence of moving breathers by stability exchange bifurcations.
Note that this does not mean that moving breathers cannot exist for
$\omega_{\mathrm{b}}>2/3$. The mechanism for the emergence of breathers when
$\omega_{\mathrm{b}}>2/3$ is simply different: it is no more than the
spontaneous motion described in Ref. \cite{Marin,Meister}. Notice that in
Hamiltonian systems, moving breathers exist over this range of frequencies
(cf. \cite{Kivshar}).

\begin{figure}[tb]
\begin{center}%
\begin{tabular}
[c]{cc}%
\includegraphics[width=4cm]{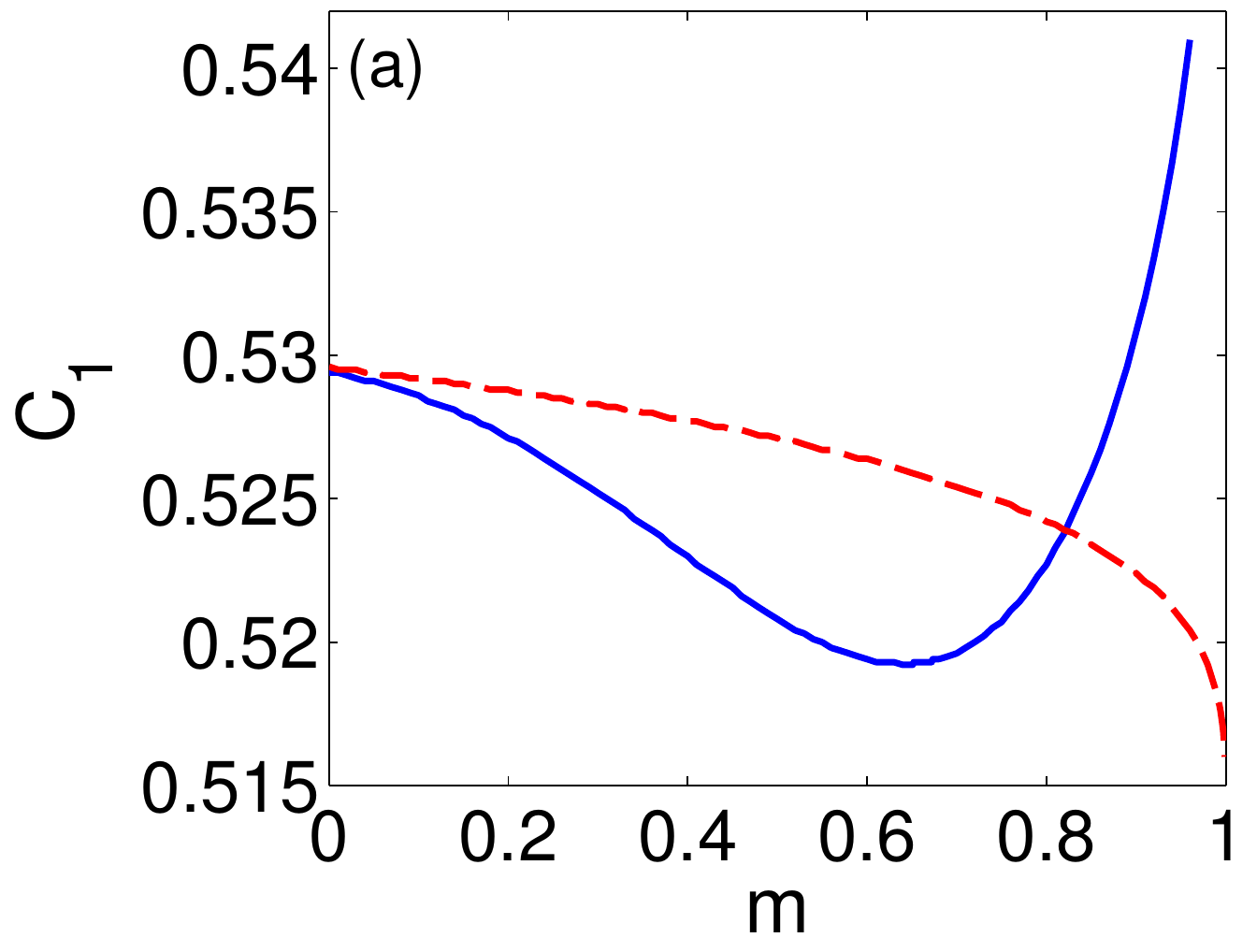} &
\includegraphics[width=4cm]{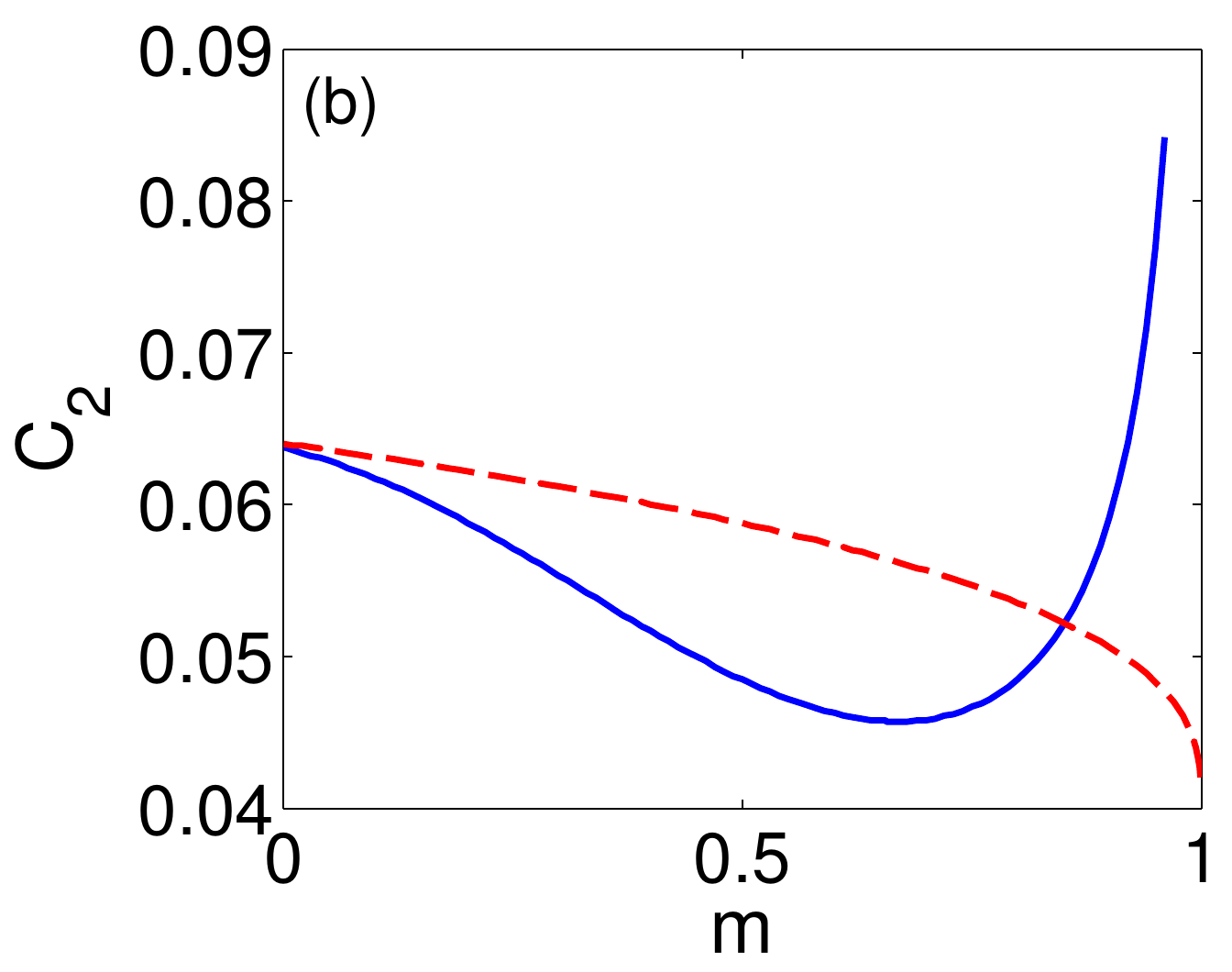}\\
&
\end{tabular}
\end{center}
\caption{(a) Loci $C_{1}$ (see the text) of the exchange of stability
bifurcation (i.e., onset of moving breathers) as a function of the shape
parameter $m$ for $\alpha=f_{0}=0.02$, $\omega_{\mathrm{b}}=0.2\pi$. (b) Loci
$C_{2}$ (see the text) of Neimark--Sacker bifurcation as a function of the
shape parameter $m$ for $\alpha=f_{0}=0.02$, $\omega_{\mathrm{b}}=0.8$. Solid
(blue) and dashed (red) lines correspond to excitations (\ref{eq:driving1})
and (\ref{eq:driving2}), respectively.}%
\label{fig:soft3}%
\end{figure}

Next, Fig. \ref{fig:soft3} shows the dependence of the critical values
$C_{1},$ $C_{2}$ as functions of the shape parameter $m$. For the excitation
$f_{1}(t)$, one sees that $C_{1}$ presents a minimum at $m\approx0.64$ when
$\omega_{\mathrm{b}}=0.2\pi$, while $C_{2}$ presents a minimum at
$m\approx0.67$ when $\omega_{\mathrm{b}}=0.8$. Notice that these values of the
shape parameter are significantly close to $0.717\approx m_{\max}$, indicating
once more again the effect of the excitation's impulse. The stability range
increases as $m$ is increased from these values (see Fig. \ref{fig:soft3}).
However, one expects that the pitchfork and Neimark-Sacker bifurcations
disappear as $m\rightarrow1$ since in such a limit the excitation and the
localization vanish. For the excitation $f_{2}(t)$, one sees that $C_{1}%
,C_{2}$ present a monotonously decreasing behavior, as expected from the
monotonously increasing behavior of the impulse $I_{2}(m)$. Thus, the analysis
of both periodic excitations $f_{1,2}(t)$ confirmed the same effect of the
excitation's impulse on the critical coupling values $C_{1,2}$.

\section{Discussion}

The numerical results discussed in Sec. \ref{sec:numerics} may be understood
by considering the excitation's impulse through an energy-based analysis,
including the properties of the action, of isolated oscillators. For the sake
of clarity, we will consider, for example, the excitation $f_{1}(t)${\textbf{
}}(\ref{eq:driving1}) in the subsequent analysis. Indeed, every breather
possesses a tail due to its localized character while the oscillators forming
this tail effectively behave as linear oscillators presenting a
small-amplitude attractor. Consequently, a breather can inherit some
properties associated with the effective linear character of the oscillators
forming its tail. We found indeed that breathers inherit the dependence on the
shape parameter according to the impulse principle. Thus, we analyze the
response of a linear (harmonic) oscillator subjected to a periodic
anti-symmetric driving:%

\begin{equation}
\label{eq:linear_osc}\ddot{u}+\alpha\dot{u}+\omega_{0}^{2}u=f_{0}\sum
_{k=0}^{\infty}G_{2k+1}\sin\left[  (2k+1)\omega_{\mathrm{b}}t\right]  ,
\end{equation}
where $G_{2k+1}$ are the Fourier coefficients of the non-harmonic excitation
(\ref{eq:driving1}):%

\begin{equation}
G_{k}=\frac{\pi^{2}N(m)k}{2\sqrt{m}K^{2}(m)}\mathrm{sech}\left[  \frac{k\pi
K(1-m)}{2K(m)}\right]  .
\end{equation}
After some straightforward algebra, one obtains the solution%

\begin{equation}
u(t)=\sum_{k=0}^{\infty}\left[  A_{2k+1}\cos\left(  (2k+1)\omega_{\mathrm{b}%
}t\right)  +B_{2k+1}\sin\left(  (2k+1)\omega_{\mathrm{b}}t\right)  \right]  ,
\end{equation}
where
\begin{align}
A_{k}  &  =f_{0}\frac{\omega_{0}^{2}-k^{2}\omega_{\mathrm{b}}^{2}}{k^{2}%
\omega^{2}\alpha^{2}+(\omega_{0}^{2}-k^{2}\omega_{\mathrm{b}}^{2})^{2}}%
G_{k},\nonumber\\
B_{k}  &  =f_{0}\frac{-k\omega_{\mathrm{b}}\alpha}{k^{2}\omega_{\mathrm{b}%
}^{2}\alpha^{2}+(\omega_{0}^{2}-k^{2}\omega_{\mathrm{b}}^{2})^{2}}G_{k}.
\end{align}
The action $J\equiv\frac{1}{2\pi}\oint\dot{u}\mathrm{d}u$ can be recast into
the form%

\begin{equation}
J=\frac{1}{\pi}\int_{0}^{T/2}\left(  \dot{u}(t)\right)  ^{2}\mathrm{d}%
t=\frac{\omega}{2}\sum_{k=0}^{\infty}k^{2}(A_{k}^{2}+B_{k}^{2}),
\end{equation}
with $T=2\pi/\omega_{\mathrm{b}}$ being the oscillator's period. Thus, the
action of the linear oscillator can be finally expressed as%

\begin{align}
\label{eq:action_linear}J  &  =\frac{\omega_{\mathrm{b}}f_{0}}{2}\sum
_{k=0}^{\infty}\mu_{2k+1}G_{2k+1}^{2},\nonumber\\
\mu_{k}  &  =k^{2}\frac{(\omega_{0}^{2}-k^{2}\omega_{\mathrm{b}}^{2}%
)^{2}+k^{2}\omega_{\mathrm{b}}^{2}\alpha^{2}}{[k^{2}\omega_{\mathrm{b}}%
^{2}\alpha^{2}+(\omega_{0}^{2}-k^{2}\omega_{\mathrm{b}}^{2})^{2}]^{2}}.
\end{align}
Notice that $\mu_{k}$ does not depend on the particular waveform of the
external periodic excitation $F(t)$, but depends on $\wb$ and
$\alpha$. Then, the dependence of the action on the shape parameter $m$
appears only in the $G_{k}^{2}$ terms. Now, after taking into account the fast
decay of the Fourier coefficients with $k$, one numerically finds that the
action presents a single maximum at $m=m_{\ell}$ which is very close to
$m_{1}$, where $m_{1}$ is the shape parameter value at which $G_{1}$ presents
a single maximum (recall from Sec. \ref{sec:numerics} that $m_{1}=0.6416$).
Notice that $m_{\ell}$ depends on $\mu_{k}$ and, consequently, on $\wb$ and $\alpha$. For instance, for the parameters sets taken in Figs.
\ref{fig:hard} and \ref{fig:soft2}, i.e. ($\alpha=0.1$, $\wb=5$) and
($\alpha=0.02$, $\wb=0.8$), the value of $m_{\ell}$ is 0.646 and
0.644, respectively.

Remarkably, the above mentioned properties also holds for the corresponding
average energies $<E>$, which for the linear oscillator reads:%

\begin{equation}
<E>=\int_{0}^{T}\left(  \frac{1}{2}\dot{u}^{2}(t)+\frac{\omega_{0}^{2}}%
{2}u^{2}(t)\right)
\end{equation}

For the excitation (\ref{eq:driving1}), it can be recast into the simple form%

\begin{align}
\label{eq:energy_linear}<E>  &  =\frac{\pi f_{0}}{2}\sum_{k=0}^{\infty}%
\mu^{\prime}_{2k+1}G_{2k+1}^{2},\nonumber\\
\mu^{\prime}_{k}  &  =\frac{(k^{2}\omega_{\mathrm{b}}^{2}+\omega_{0}%
^{2})[(\omega_{0}^{2}-k^{2}\omega_{\mathrm{b}}^{2})^{2}+k^{2}\omega
_{\mathrm{b}}^{2}\alpha^{2}]}{\omega_{\mathrm{b}}[k^{2}\omega_{\mathrm{b}}%
^{2}\alpha^{2}+(\omega_{0}^{2}-k^{2}\omega_{\mathrm{b}}^{2})^{2}]^{2}}.
\end{align}

\begin{figure}[tb]
\begin{center}%
\begin{tabular}
[c]{cc}%
\includegraphics[width=4cm]{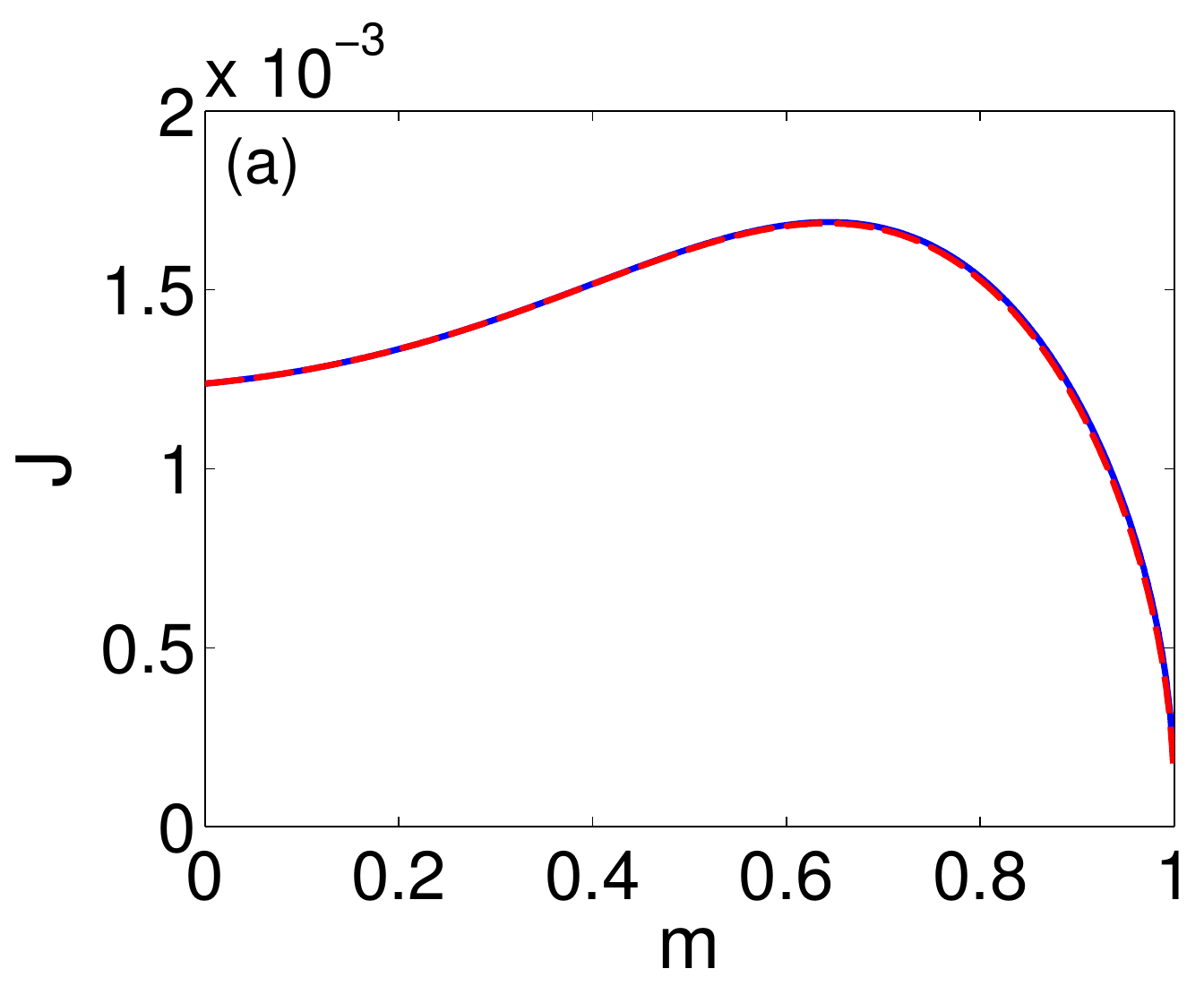} &
\includegraphics[width=4cm]{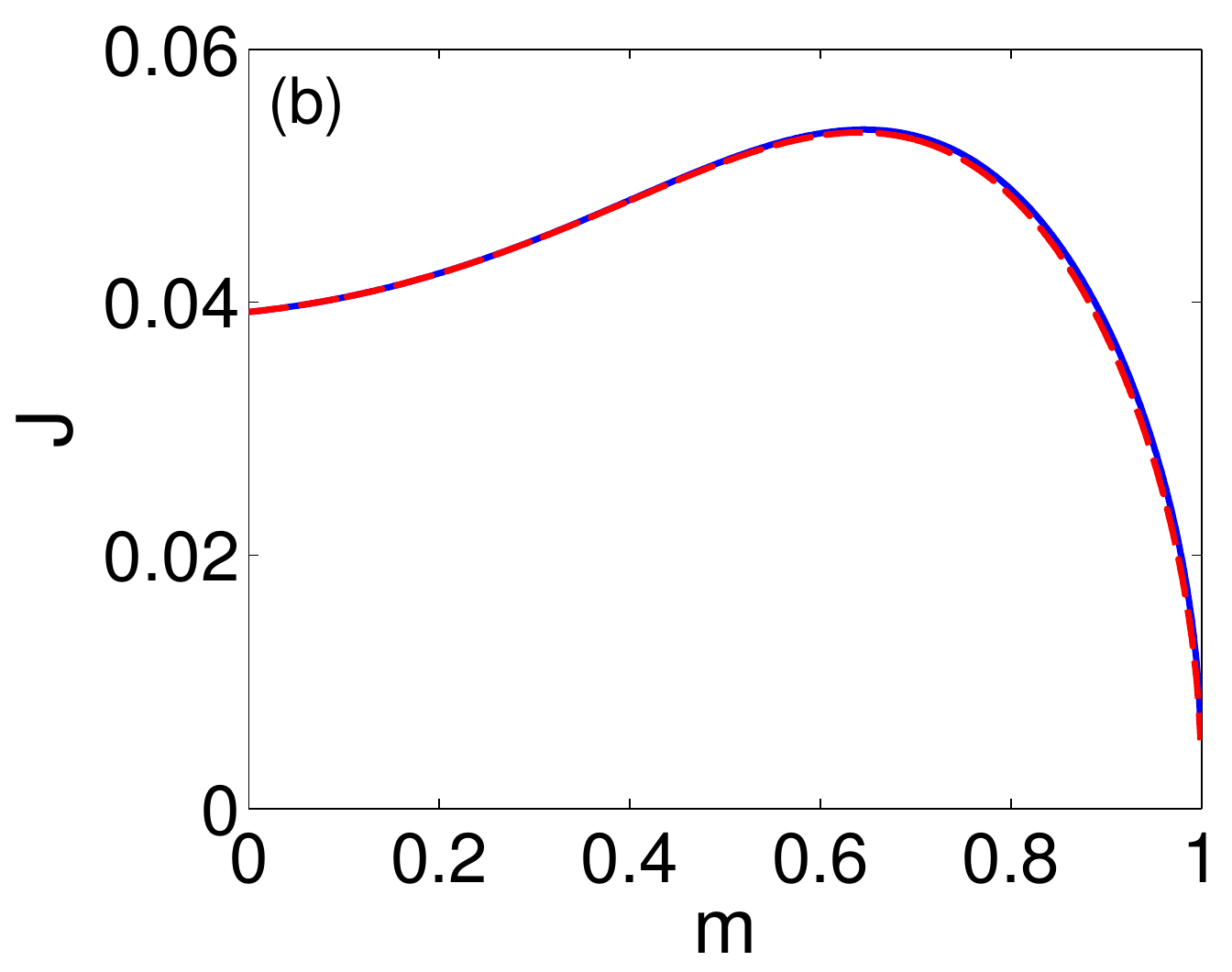}\\
\includegraphics[width=4cm]{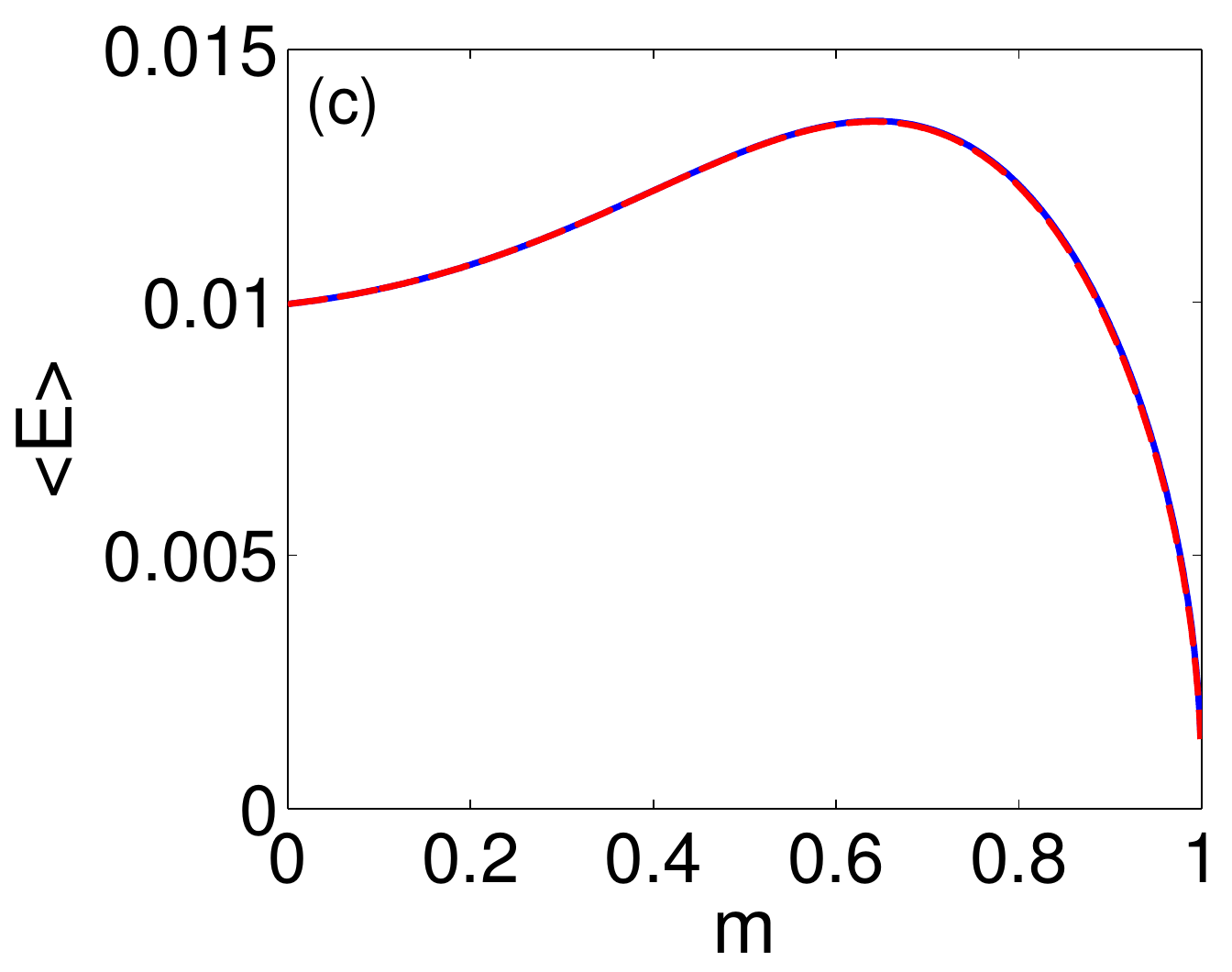} &
\includegraphics[width=4cm]{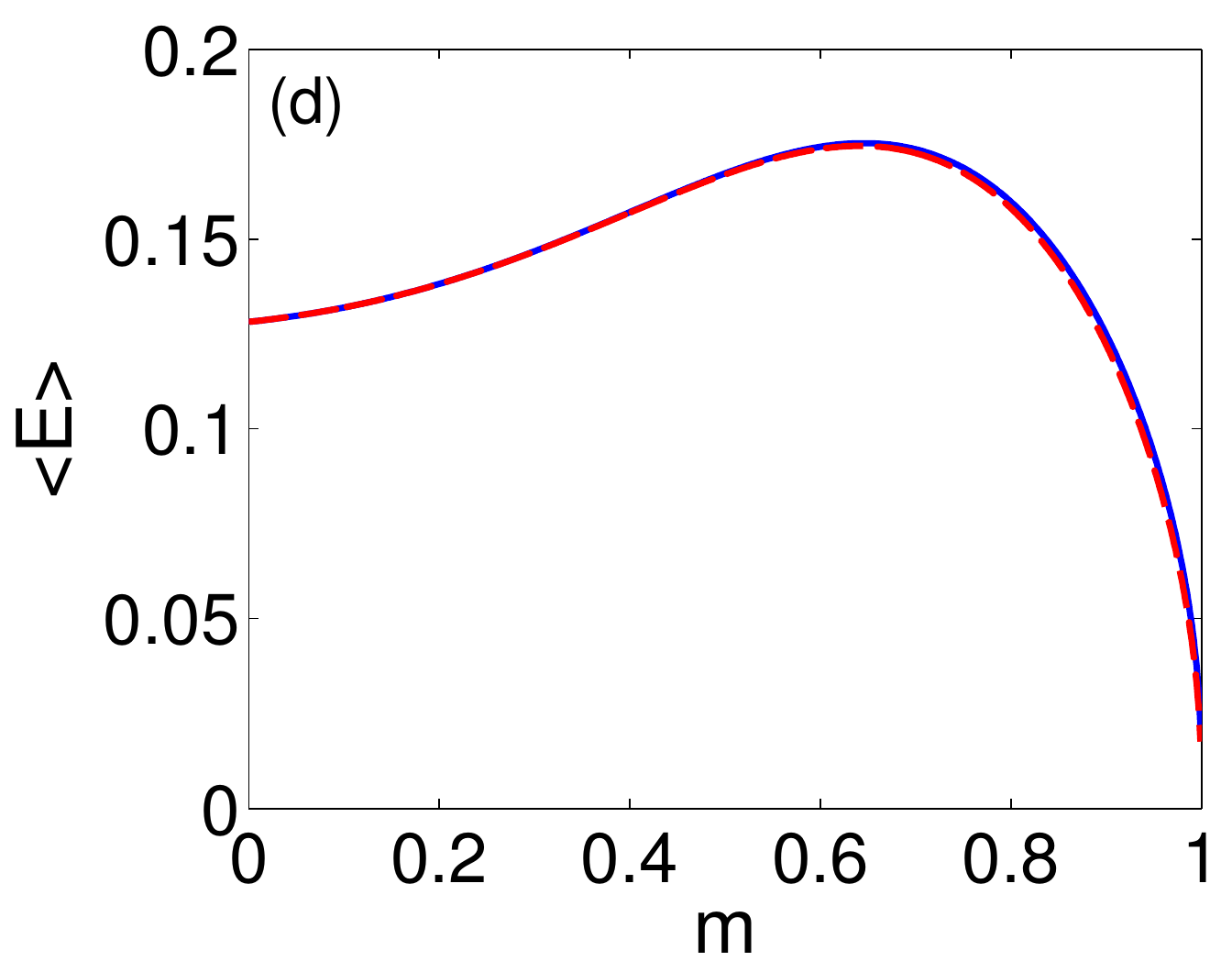}\\
&
\end{tabular}
\end{center}
\caption{Action $J$ [Eq. (\ref{eq:action_linear})] (top panels) and average
energy $<E>$ [Eq. (\ref{eq:energy_linear})] (bottom panels) of solutions of
the linear oscillator [Eq. (\ref{eq:linear_osc})] vs shape parameter $m$ for
the complete Fourier series (solid lines) of excitation (3) and the main
harmonic approximation ($G_{2k+1}=0\ \forall k\geq1$; dashed lines) and two
sets of parameters: (left panels) $\omega_{\mathrm{b}}=0.8$, $f_{0}=0.02$,
$\alpha=0.02$. (right panels) $\omega_{\mathrm{b}}=5$, $f_{0}=3$, $\alpha
=0.1$.}%
\label{fig:action1}%
\end{figure}

Figure \ref{fig:action1} shows the dependence of the action and average energy
of solutions of the linear oscillator [Eq. (\ref{eq:linear_osc})] on the shape
parameter for the complete Fourier series and the main harmonic approximation
($G_{2k+1}=0,\ \forall k\geq1$) and two sets of the remaining parameters.

As already anticipated in Sec. \ref{sec:intro}, threshold phenomena associated
with breathers' emergence and stability exhibit a high sensitivity to the
excitation's impulse. To show this, we start with a general argument showing
the relationship between energy changes and the quantities action and impulse
for periodic solutions of isolated (nonlinear) oscillators. After integrating
the corresponding energy equation over half a period (see, e.g., \cite{6,7}),
one obtains%

\begin{equation}
\Delta E\equiv E(T/2)-E(0)=-\alpha\int_{0}^{T/2}\left(  \dot{u}^{2}(t)\right)
\mathrm{d}t+\int_{0}^{T/2}\dot{u}(t)F(t)\mathrm{d}t.
\end{equation}
Now, after applying the first mean value theorem for integrals \cite{GR} to
the last integral on the right-hand side of Eq. (\ref{eq:energy_linear}) and
recalling the definitions of action and impulse, one obtains%

\begin{equation}
\Delta E=-\alpha\pi J+T\dot{u}(t^{\ast})I,
\end{equation}
where $t^{\ast}\in\lbrack0,T/2]$ while $J$ and $I$ are the action and the
impulse, respectively. Note that $t^{\ast}$ becomes independent of the
excitation's waveform as $T\rightarrow0$ \cite{6,7}. It should be stressed
that this limiting regime is unreachable for the present case of discrete
breathers in nonlinear chains, specially in the soft potential case, due to
breather frequencies are always below a maximum, and hence they cannot be
increased arbitrarily. Therefore, we see that the dependence of $\Delta E$ on
$m$ for any (linear or nonlinear) isolated oscillator relies on the dependence
on $m$ of $J$ and $\dot{u}(t^{\ast})I$. For the case of a linear oscillator
and the excitation $f_{1}(t)${\textbf{ }}(\ref{eq:driving1}), one readily
obtains that $\dot{u}(t^{\ast})I$ can be expressed as $\sum_{k}\mu
_{2k+1}^{(1)}G_{2k+1}^{2}$, where $\mu_{k}^{(1)}$ is independent on $m$, and
hence the energy will present a maximum at $m=m_{\ell}^{\prime}\approx
m_{\ell}$. We find that the dependence of the action on the shape parameter
for nonlinear oscillators is quite similar to that of the discussed linear case.

\begin{figure}[tb]
\begin{center}%
\begin{tabular}
[c]{cc}%
\includegraphics[width=4cm]{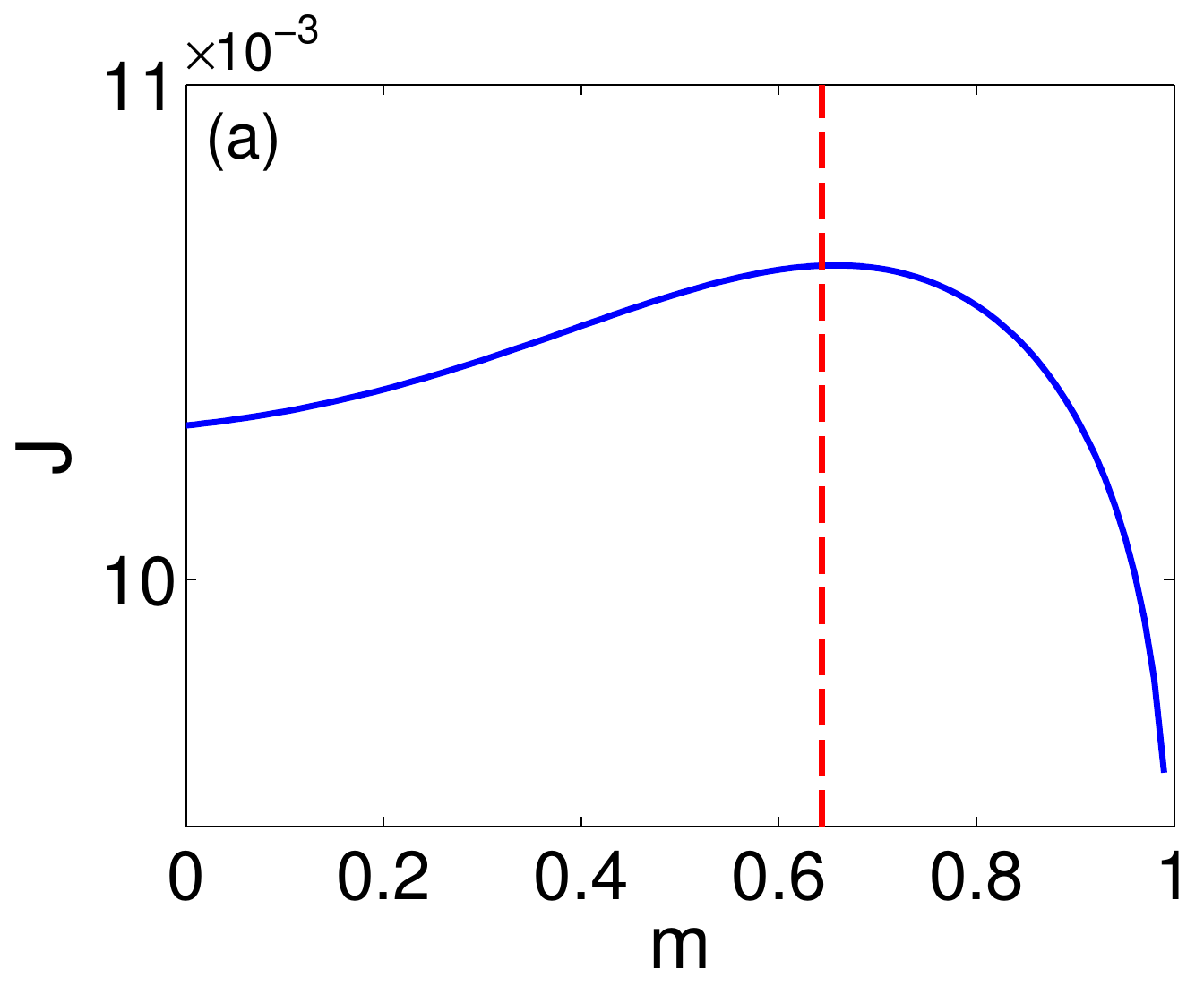} &
\includegraphics[width=4cm]{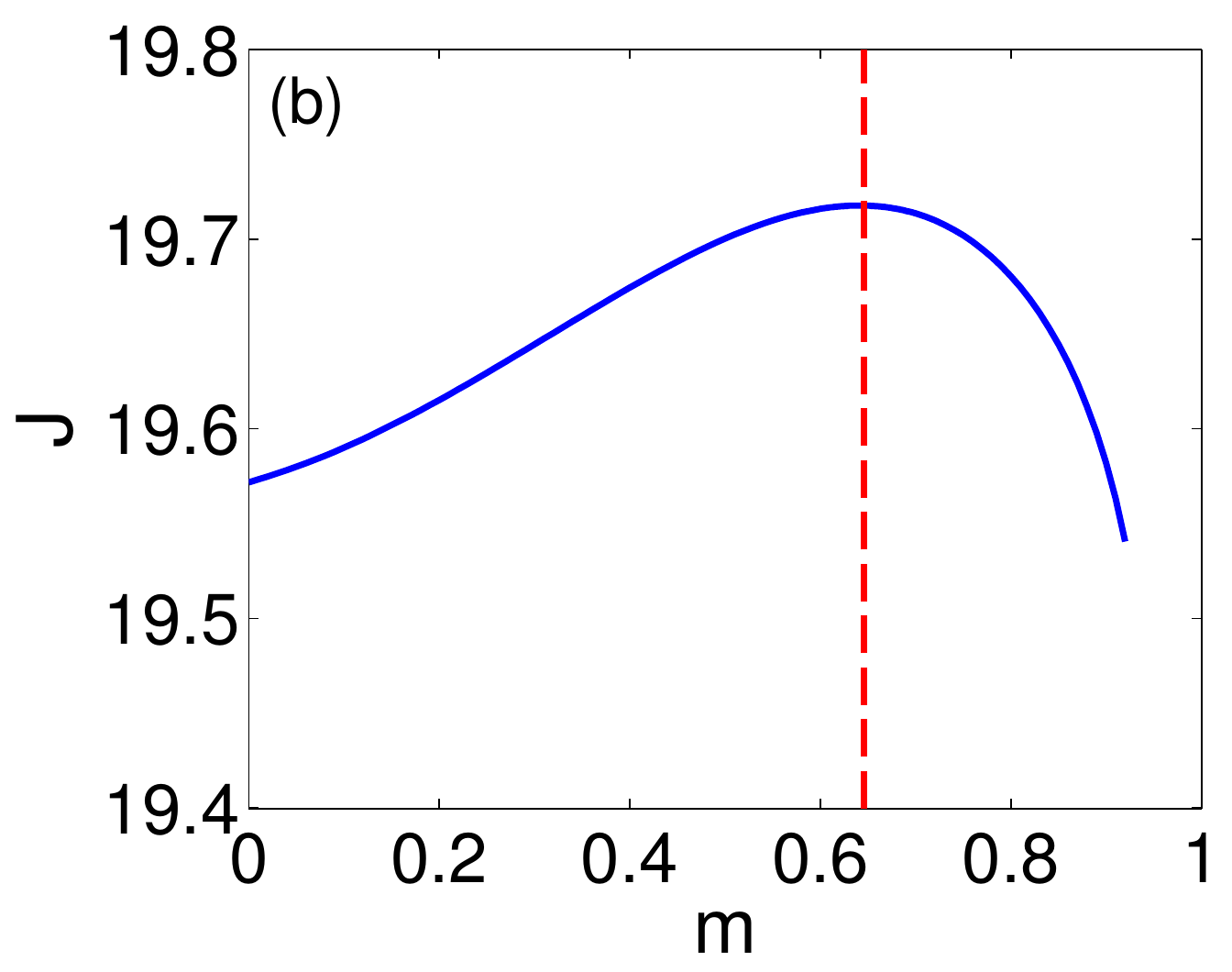}\\
\includegraphics[width=4cm]{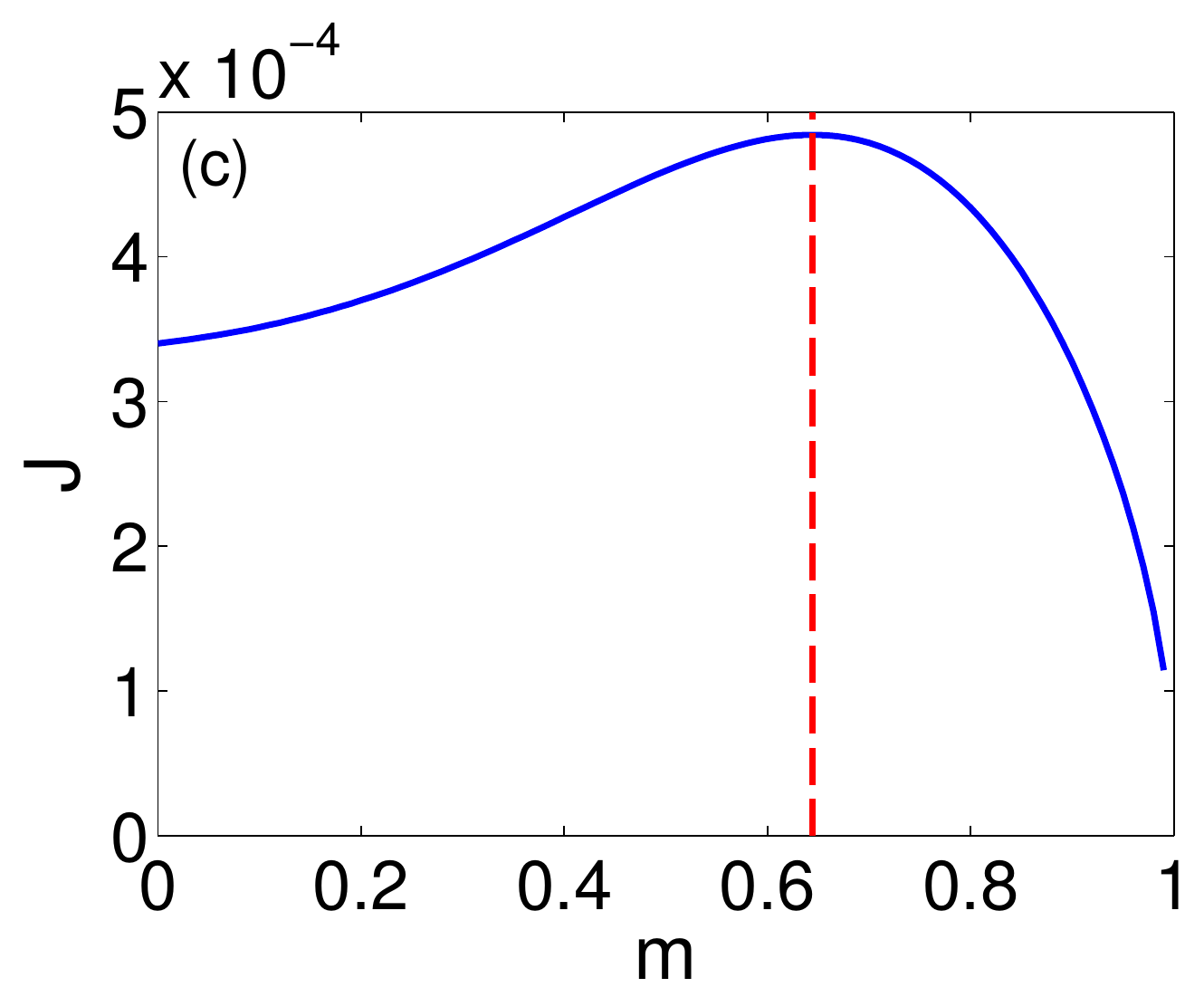} &
\includegraphics[width=4cm]{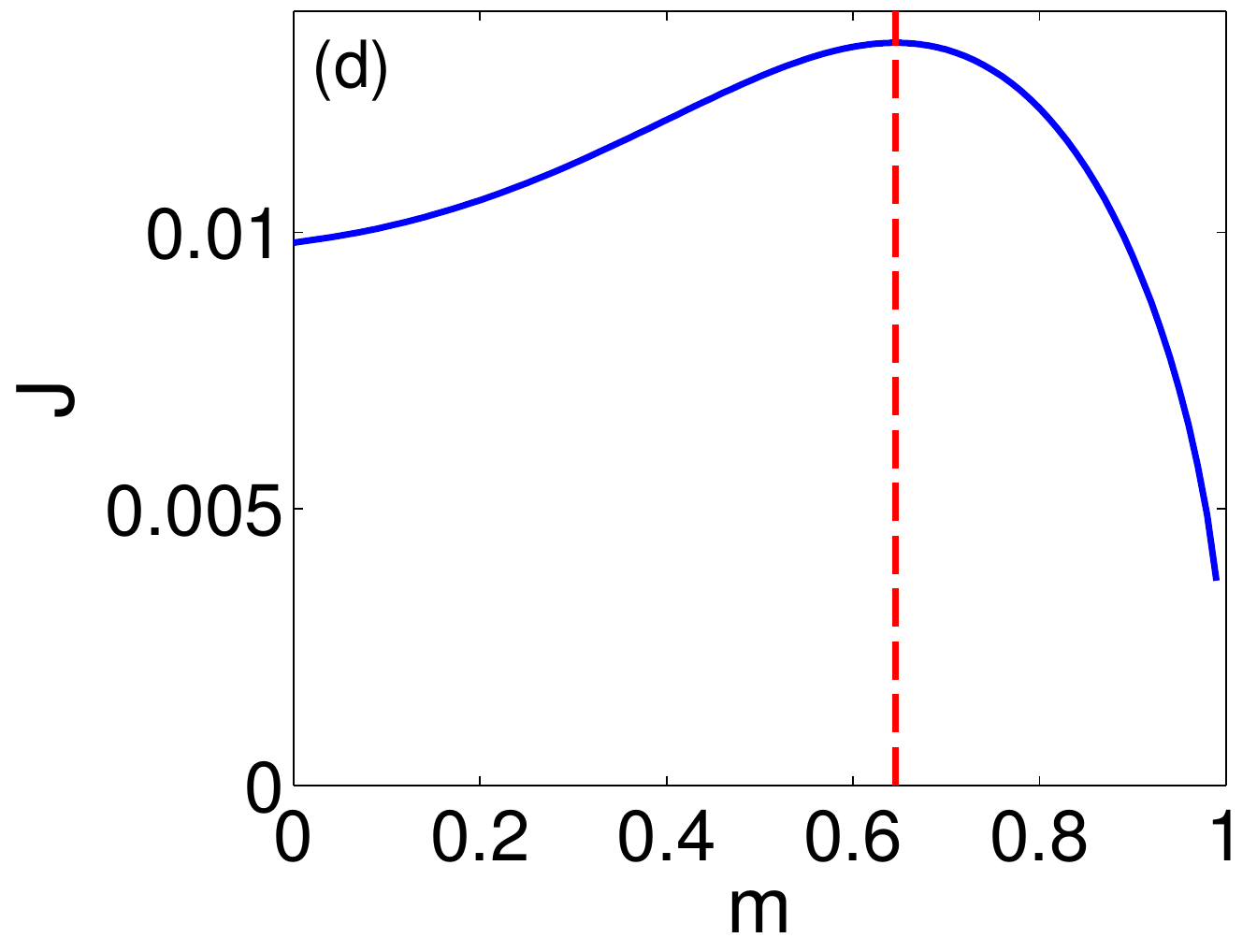}\\
&
\end{tabular}
\end{center}
\caption{Action $J$ of (top panels) large-amplitude attractors and (bottom
panels) small-amplitude attractors corresponding to an isolated nonlinear
oscillator with (left panels) a sine-Gordon potential for the excitation (3),
$\omega_{\mathrm{b}}=0.8$, $f_{0}=0.02$, $\alpha=0.02$, and (right panels) a
$\phi^{4}$ potential for $\omega_{\mathrm{b}}=5$, $f_{0}=3$, $\alpha=0.1$. The
vertical dashed lines indicate the $m$ value where the action of the linear
oscillator is maximum, i.e. $m=m_{\ell}$ (cf. Fig. \ref{fig:action1}).}%
\label{fig:action2}%
\end{figure}

\begin{figure}[tb]
\begin{center}%
\begin{tabular}
[c]{cc}%
\includegraphics[width=4cm]{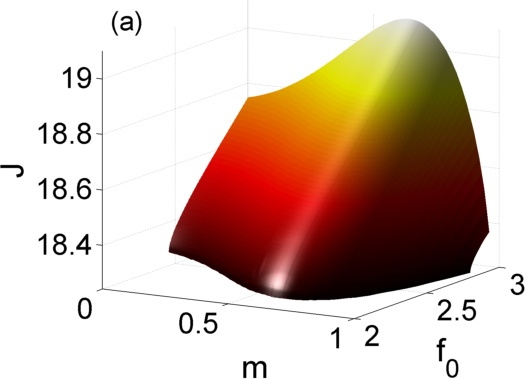} &
\includegraphics[width=4cm]{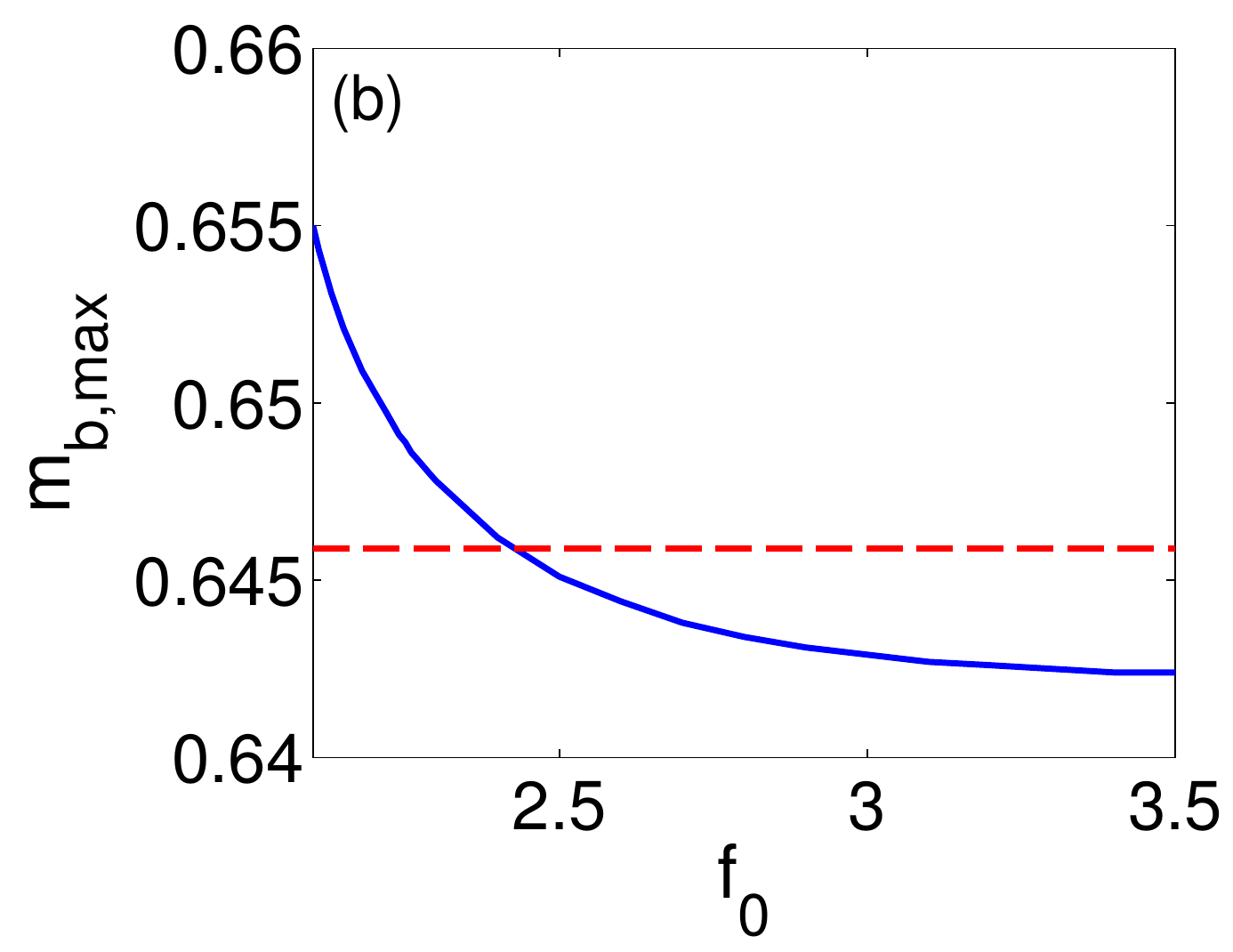}\\
&
\end{tabular}
\end{center}
\caption{(a) Action $J$ of a breather emerging in a $\phi^{4}$ chain vs
$f_{0}$ and $m$ in the range $f_{0}\in(2.06,3)$ for the excitation (3). (b)
Shape parameter value at which the action is maximum $m_{b,\max}$ vs amplitude
$f_{0}$; horizontal line corresponds to $m=m_{\ell}=0.646$. Fixed parameters:
$\omega_{\mathrm{b}}=5$, $\alpha=0.1,$ $C=1$.}%
\label{fig:action_breather}%
\end{figure}

To connect this analysis of isolated oscillators with discrete breathers of
nonlinear chains (\ref{eq:dyn}), one has to calculate the action of a
breather, $J=\sum_{n}\oint\dot{u}_{n}\mathrm{d}u_{n}$, on the one hand, and to
distinguish between the periodic attractors with large and the small
amplitudes, on the other hand, since the orbits associated with the latter can
substantially differ from those of a strictly linear oscillator despite of its
relatively small oscillation amplitude. In any case, numerical simulations
confirmed that the value $m=m_{\max}\approx0.717$ at which the impulse
function $I_{1}(m)$ presents a single maximum is very close to $m_{\ell}$ in
the sense that the waveforms corresponding to $m_{\max}$ and $m_{\ell}$ (and
$m_{1}$) can be hardly distinguishable. Figure \ref{fig:action2} shows an
illustrative example for the cases of a hard $\phi^{4}$ potential and a
sine-Gordon potential, while Fig. \ref{fig:action_breather} shows, for the
case of a hard $\phi^{4}$ potential, that the breather action presents a
single maximum at $m=m_{b,\max}$ which is also very close to $m_{\ell}%
\approx0.646$.

\section{Conclusions}

We have shown through the example of a discrete nonlinear Klein-Gordon
equation that varying the impulse transmitted by periodic external excitations
is a universal procedure to reliably control the generation of stationary and
moving discrete breathers in driven dissipative chains capable of presenting
these intrinsic localized modes. We have analytically demonstrated that the
enhancer effect of the excitation's impulse, in the sense of facilitating the
generation of stationary and moving breathers, is due to a correlative
increase of the breather's action, while numerical experiments corresponding
to the cases of a hard $\phi^{4}$ potential and a sine-Gordon potential
confirmed the effectiveness of the impulse\textbf{ }as the relevant quantity
controlling the effect of the external excitation. The consideration of this
relevant quantity opens up new avenues for studying
external-excitation-induced phenomena involving intrinsic localized modes in
discrete nonlinear systems, including, for instance, breather-to-soliton
transitions and emergence of chaotic breathers. Our present work is aimed to
explore these and related problems.

\begin{acknowledgments}
R. C. gratefully acknowledges financial support from the Junta de Extremadura
(JEx, Spain) through Project No. GR15146.
\end{acknowledgments}

\end{document}